\documentclass[ superscriptaddress,amsmath, amssymb, aps, prd, twocolumn, lengthcheck, sort&compress, showpacs,  nofootinbib, 10pt,natbib ]{revtex4-1}
\usepackage{graphicx}
\usepackage{dcolumn}
\usepackage{bm}
\usepackage{graphicx}
\usepackage{epsfig}
\usepackage{dcolumn}
\usepackage{subfigure}
\usepackage[usenames]{color}
\usepackage[colorlinks=true,linkcolor=blue,citecolor=blue,urlcolor=blue]{hyperref}
\usepackage{textfit}
\usepackage{bbm}
\usepackage{natbib}

\allowdisplaybreaks[1]
\definecolor{mycolor}{rgb}{0.2, 0.7, 0.8}

\newcommand{\pslash}{\not\!p}
\newcommand{\qslash}{\not\!q}
\newcommand{\kslash}{\not\!k}

\newcommand{\kizilersu}{K{\i}z{\i}lers{\" u}}

\begin{document}

\preprint{JLAB-THY-14-1935}

\title{Dynamical mass generation in unquenched QED using the Dyson--Schwinger equations}
\author{ Ay{\c s}e K{\i}z{\i}lers{\" u} }
  \email{akiziler@physics.adelaide.edu.au}
  \affiliation{Special Research Centre for the Subatomic Structure of Matter, School of Chemistry and Physics, Adelaide University, 5005, Australia}
\author{Michael R.\ Pennington}
  \email{michaelp@jlab.org}
  \affiliation{Thomas Jefferson National Accelerator Facility, Newport News, Virginia 23606, USA}
\author{Tom Sizer}
  \email{tsizer@physics.adelaide.edu.au}
  \affiliation{Special Research Centre for the Subatomic Structure of Matter, School of Chemistry and Physics, Adelaide University, 5005, Australia}
\author{Anthony G.\ Williams}
  \email{anthony.williams@adelaide.edu.au}
  \affiliation{Special Research Centre for the Subatomic Structure of Matter, School of Chemistry and Physics, Adelaide University, 5005, Australia}
\affiliation{ARC Centre of Excellence for Particle Physics at the Tera-scale, School of Chemistry and Physics, Adelaide University, 5005, Australia}
\author{Richard Williams}
\email{richard.williams@theo.physik.uni-giessen.de} 
\affiliation{Institut f\"ur Theoretische Physik, Justus-Liebig--Universit\"at Gie{\ss}en, 35392 Gie{\ss}en, Germany}

\date{\today}

\begin{abstract}
\noindent
We present a comprehensive numerical study of dynamical mass generation for unquenched QED in
four dimensions, in the absence of four-fermion interactions, using the Dyson--Schwinger approach.
We begin with an overview of previous investigations of criticality in the quenched approximation.
To this we add an analysis using a new fermion-antifermion-boson interaction ansatz, the \kizilersu-Pennington (KP)
vertex, developed for an unquenched treatment. After surveying criticality in previous unquenched studies,
we investigate the performance of the KP vertex in dynamical mass generation using a renormalized
fully unquenched system of equations.
 This we compare with the results for two hybrid vertices incorporating the Curtis--Pennington vertex in the fermion equation.
We conclude that the KP vertex is as yet incomplete, and its relative gauge-variance
is due to its lack of massive transverse components in its design.
\end{abstract}

\pacs{11.15.Ex,12.20.-m,12.38.Cy}
 \maketitle

\section{\label{sec:Introduction}Introduction}

Quantum Electrodynamics has served as a prototype field theory for studying both perturbative and non-perturbative phenomena for many years.
Though the coupling constant in nature is small, and hence  perturbative expansions meaningful, one can
imagine a theory of strongly-coupled QED where such an approach is inappropriate. The attraction of looking at such a
scenario lies with its relatively simple Abelian gauge structure and fermion-antifermion-photon interaction: QCD, by way of contrast, although naturally exhibiting a
strong-coupling regime, is non-Abelian and requires knowledge of the quark-gluon~\cite{Alkofer:2008tt,Williams:2014iea,Aguilar:2014lha} and three-gluon vertices~\cite{Aguilar:2013vaa,Blum:2014gna,Eichmann:2014xya}.

Essentially, there are two mainstream approaches to non perturbative
studies of QED: those on the lattice and those using a continuum approach such as the Dyson--Schwinger Equations. Finite volume
calculations on the lattice tend to concentrate on the three-dimensional variant, where the phase structure of the theory is
of interest not only because of  its analogies to QCD, but also for its potential application to the study of cuprate superconductors. Functional methods have found that finite volume effects are large~\cite{Gusynin:2003ww,Goecke:2008zh}, which is significant for more realistic models featuring anisotropy~\cite{Franz:2002qy,Tesanovic:2002zz,Bonnet:2011hh}. 
Though lattice studies have looked at the four dimensional case, there has been little progress over the
last few years due to complications associated with the four-fermion operator. This causes difficulties when comparing lattice studies with pure QED as calculated using the continuum approach.

A key component in developing our understanding of non-perturbative physics has been the study of the mechanism of dynamical mass generation. The DSE approach to this is to calculate the fermion and photon propagators numerically.
This requires knowledge of the fermion-photon vertex, which can either be provided by Ansatz or calculated from its 
DSE. In the absence of a bare mass, it is universally found that the fermion mass function is non-zero only above some critical value of the coupling.
The majority of these studies have been limited to the quenched theories \cite{Fomin:1976af,Miransky:1984ef,Miransky:1986xp,Miransky:1986ib,Miransky:1979ks,Fukuda:1976zb,Fomin:1984tv,Lombardo:1994vz,Atkinson:1990bg,Atkinson:1986aw,Atkinson:1993mz,Curtis:1993py,Curtis:1992jg,Kizilersu:2000qd,Kizilersu:2001,Bloch:1994if,Bloch:1995dd,Williams:2007zzh,Bashir:1994az,Bashir:2011dp,Gusynin:1998se,Hawes:1991,Hawes:1994ce,Hawes:1996mw,Williams:1995rd,Reenders:1999fz,Fomin:1984fd,Roberts:1994dr} where the photon propagator is tree-level and the coupling does not run:
only few works have been devoted to the unquenched (complete) theory \cite{Bloch:1994if,Bloch:1995dd,Gusynin:1989mc,Gusynin:1989ca,Kondo:1990ig,Kondo:1990ky,Kondo:1990ar,Kondo:1990st,Oliensis:1990sg,Ukita:1990uw,Kondo:1991ms,Williams:2007zzh,Bashir:2011ij,Akram:2012jq,Kizilersu:2013hea}. This has historically involved two limitations - the computational challenge recently ameliorated by the emergence of faster
computers, and the incomplete knowledge of the explicit form of the 3-point fermion-photon Green's function (the vertex). Today, progress has been made towards directly solving the vertex DSEs, with most of the attention focused upon QCD~\cite{Maris:1999bh,Kellermann:2008iw,Huber:2012zj,Blum:2014gna,Eichmann:2014xya,Williams:2014iea}. However, complementary to this is the explicit construction of vertex models constrained by consideration of functional identities~\cite{Aguilar:2013vaa,Rojas:2013tza,Aguilar:2014lha}.

In this paper we present a comprehensive study of dynamical mass generation in strong coupling four dimensional QED
 using the DSEs. The numerical analysis has been performed independently by two groups, MRP and RW~\cite{Williams:2007zzh}, and AK, TS and AGW~\cite{Kizilersu:2013hea, Sizer:phd}. We report jointly upon a recent Ansatz~\cite{Kizilersu:2009kg}, dubbed the \kizilersu--Pennington vertex (KP), and explore its properties as a function of coupling strength, gauge parameter, and fermion number. The paper is organized as follows. In section \ref{sec:Formulation} the DSE formalism is introduced, followed by a discussion of multiplicative renormalizability and its importance, section~\ref{sec:MR}. Two fermion-photon vertices constrained by this are presented, that of Curtis--Penningtion (CP) vertex for quenched QED~\cite{Curtis:1990zs} and the KP vertex for unquenched QCD. In section  \ref{sec:QUENCHED QED} we give a numerical survey of the critical fermion number in quenched massless QED$_{4}$ and compare to the KP vertex. Strictly massless solutions with the unquenched KP vertex are presented
 in section \ref{sec:Massless Unquenched QED}. Fermion flavour criticality and dynamical mass generation for a variety of gauges is discussed in section \ref{sec:Massive Unquenched QED and Dynamical Mass Generation}, where hybrid vertex models that incorporate the CP vertex in the fermion equation are also studied. We conclude in section \ref{sec:Conclusion}.

\section{\label{sec:Formulation} Dyson--Schwinger Equations}

\begin{figure}[t]
\begin{center}
\includegraphics[scale = 0.65]{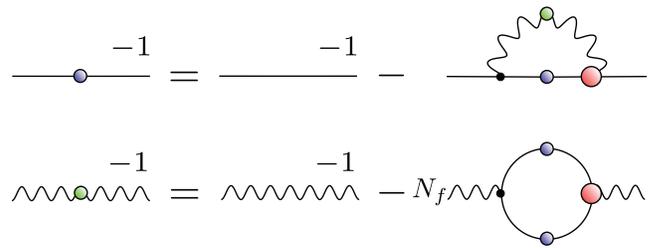}\\
\end{center}
\caption{The Dyson--Schwinger Equations in QED for the fermion and photon propagators (straight and wavy lines respectively). Full circles denote fully dressed quantities.} 
\label{fig:DSE}
\end{figure}

In a Quantum Field Theory, all fundamental quantities may be related to the underlying Green's functions that describe
the theory. In Euclidean space, where with almost no exception non-perturbative calculations are performed, the
equations of motion describing these correlation functions are the Dyson--Schwinger Equations (DSEs). The lowest
order DSEs relevant to QED are shown in Fig.~\ref{fig:DSE}. These are the first of an infinite tower of coupled non-linear integral equations, relating Green's functions of different orders.  Each of these with $N_F$ fermion legs and $N_A$ photon legs satisfies its own equation. These couple each one-particle irreducible Green's function to others as illustrated in Fig.~\ref{fig:DSE} and take the form of non-linear integral equations. The DSEs for two point functions (the fermion and photon propagators) are shown in 
Fig.~\ref{fig:DSE} and given explicitly in Eqs.~(\ref{eq:mainsdf}, \ref{eq:mainsdph}) to come.
To solve the infinite tower is impossible, consequently, some form of truncation must be introduced. Minimally, to evaluate the two point functions, we need to introduce a suitable Ansatz for the fermion-photon vertex that appears in both of  the equations in Fig.~\ref{fig:DSE}. To be realistic, such an Ansatz must attempt to encode the effect of all the higher point Green's functions, at least as far as their implication for the fermion and photon propagators is concerned.  The treatment discussed in section \ref{sec:MR} is an example of this.

\subsection{\label{sec:Fermion-Photon Propagators}Fermion and Photon Propagators }

The renormalized DSEs for the fermion and photon propagators are
\begin{align}
S^{-1}(p) &= Z_2\,{S^{(0)}}^{-1}(p) \; \nonumber\\
          &- \frac{Z_2^2}{Z_1}\;i e^2 \int^\Lambda_M \widetilde{dk}\;
           \Gamma^\mu (p,k;q) \; S(k)\;\gamma^\nu \; \Delta_{\mu \nu}(q)\;,
\label{eq:mainsdf} \\
\Delta_{\mu \nu}^{-1}(q) &= Z_3\;{\Delta_{\mu \nu}^{(0)}}^{-1}(q) \> \nonumber\\
                          &+ \frac{Z_2^2}{Z_1}\;i e^2 N_F \; {\rm Tr} \int^\Lambda_M \widetilde{dk}\; 
                          \Gamma_\mu (p,k;q)\;S(k) \;\gamma_\nu \;  S(p)\; .
\label{eq:mainsdph} 
 \end{align}
\noindent
Here $\widetilde{dk}=d^4k/\left(2\pi\right)^4$, $q = k-p$, $S^{(0)}$ and $\Delta_{\mu \nu}^{(0)}$ are the tree-level fermion and photon propagators respectively, and 
the $Z_{i}$ factors relate the unrenormalized quantities arising from the Lagrangian to the corresponding
renormalized quantities appearing in these equations.
Renormalization allows us to trade the momentum cut-off $\Lambda$ for some physical renormalization
point $\mu$, so $Z_{i} = Z_i(\mu^2,\Lambda^2)$.

Explicitly, these relations are
\begin{align}
\Gamma_\mu(k,p;\mu) \, =\, Z_1\,\Gamma^0_\mu(k,p;\Lambda)
\end{align}
for the fermion-photon vertex,
\begin{align}
\label{eq:renfermion}
S(p;\mu) &= Z_2^{-1}\,S^0(p;\Lambda) \,,\\
D_{\mu\nu}(q;\mu) &= Z_3^{-1}\,D^0_{\mu\nu}(q;\Lambda)\;,
\label{eq:renphoton}
\end{align}
for the fermion and photon propagators, and
\begin{align}
\alpha(\mu) &= (Z_2/Z_1)^2\,Z_3\,\alpha_0\;,
                        \label{eq:renalpha1}
\end{align}
for the renormalized coupling strength,
where $\alpha_0 = e_{0}^2/4\pi$ and $\alpha = e^2/4\pi$, and unrenormalized quantities in the preceding paragraph are indicated by a $0$ superscript or subscript.

As is well-known, the Dirac structure of the fermion propagator can be decomposed as
\begin{align}\label{FermionPropagator}
    S(p) = \frac{F(p^2)}{\pslash - M(p^2)} = \frac{1}{A(p^2)\pslash - B(p^2)}\,,
\end{align}
which implies
\begin{align}
    F(p^2) = \frac{1}{A(p^2)}\,,\quad
    M(p^2) = \frac{B(p^2)}{A(p^2)}\,.
\end{align}
While the fermion propagator comprises two gauge-dependent scalar functions, $F(p^2)$, the fermion wave-function renormalization function,
and $M(p^2)$, the mass function,
the full photon propagator is characterised by one gauge-independent function, $G(p^2)$, the photon wave function renormalization,
\begin{align}
\label{PhotonPropagator}
    \Delta_{\mu\nu}(q)
        = \frac{-1}{q^2}
            \left[
                G(q^2)\left(g_{\mu\nu}-\frac{q_{\mu} q_\nu}{q^2}\right)+
                \xi\frac{q_\mu q_\nu}{q^2}
            \right]\,,
\end{align}
where $\xi$ is the covariant gauge parameter.
The gauge parameter is also renormalized via $\xi_0=Z_\xi\,\xi$, from which the invariance of 
$\alpha\xi$ implies $Z_\xi=Z_3$.

The respective tree-level propagators are obtained by setting 
$A= G =1$ and $B = m_0$ in the above, with $m_0$ the bare fermion mass appearing in the Lagrangian.

In this Abelian theory, the fermion-photon
vertex must satisfy the Ward-Green-Takahashi identity (WGTI)~\cite{Ward,Green:1953te,Takahashi}: 
\begin{align}
\label{eq:vertexWTI}
    Z_1\,q_\mu \, \Gamma^\mu(p,k) = Z_1\, q_\mu \, \Gamma_L^\mu(p,k) = Z_2\,S^{-1}(k) - Z_2\,S^{-1}(p) \,.
\end{align}
This plays a role in constraining the explicitly gauge-dependent part of the fermion DSE.
The corresponding equation for the unrenormalized quantities results in the identification
\begin{align}
	Z_1 = Z_2\;,
\end{align}
so Eq.~(\ref{eq:renalpha1}) simplifies to
\begin{align}
	\alpha(\mu^2) &=Z_3\,\alpha_0\,.
                        \label{eq:renalpha}
\end{align}

The renormalization point invariant running coupling is then given by
\begin{align}
\alpha(p^2) = \alpha(\mu^2)\, G(p^2,\mu^2)\, ,
\end{align}
with $G$ the photon dressing function. The mass function $M(p^{2})$ is also a renormalization point invariant.
An important consequence of the gauge symmetry is that the photon dressing function $G$ is independent of the gauge parameter $\xi$,
which itself receives no higher-order corrections.

\subsection{\label{sec:Fermion-Photon Non-perturbative vertices} Fermion-Photon Vertex}

The fermion-photon vertex consists of twelve spin amplitudes built from $I, \gamma^{\mu}$ and two
independent four-momenta, $k^{\mu}$ and $p^{\mu}$. These can be combined into four longitudinal components, $L_i$, whose coefficient
functions $\lambda_i$  are wholly determined by the WGTI as Ball and Chiu stated \cite{Ball:1980ay},
 and eight transverse components $T_i$ satisfying
\begin{align}
	q_{\mu}\,T_i^{\mu}(p,k) = 0\;; \quad T_i^{\mu}(k,k) = 0\;,
\end{align}
(that is, they are orthogonal to the photon momentum and free of kinematic singularities) such that
\begin{align}
    \Gamma^\mu(p,k) = \sum_{i=1}^4 \lambda_i \, L_i^\mu(p,k) +  \sum_{i=1}^8 \tau_i \, T_i^\mu(p,k)\;,
\label{eq:transverse}
\end{align}
where $\lambda_i=\lambda_i(p^2,k^2,q^2)$ and $\tau_i=\tau_i(p^2,k^2,q^2)$.
We employ the same basis for $L_i$ as in \cite{Kizilersu:1995iz,Kizilersu:2009kg}~: 
\begin{align}
        L_1(p,k) &=\gamma^\mu \,,\nonumber\\
        L_2(p,k) &= (\kslash + \pslash) (k+p)^\mu \,,\nonumber\\
        L_3(p,k) &= (k+p)^\mu \,, \nonumber\\
        L_4(p,k) &= (k^{\nu}+p^{\nu})\sigma^{\mu\nu}\,,
\label{eq:Ls}
\end{align}
\noindent
with $\sigma_{\mu\nu}=\frac{1}{2}\,[\gamma_\mu,\gamma_\nu]$.
\noindent
The longitudinal components are fixed uniquely by the WGTI to be (superscript `M' denotes Minkowski space)
\begin{align}
    \lambda^M_1(p^2,k^2) &=
                   \frac{1}{2}          \left[A_k + A_p \right]\,, \nonumber\\
    \lambda^M_2(p^2,k^2) &=
                   \frac{1}{2}\frac{1}{(k^2-p^2)} \left[A_k - A_p \right] \,,\nonumber\\
    \lambda^M_3(p^2,k^2) &=
                   -\frac{1}{k^2-p^2}   \left[B_k - B_p \right]\,, \nonumber\\[2mm]
    \lambda^M_4(p^2,k^2) &= 0\,,
\end{align}
where $A_k = A(k^2)$ is a convenient shorthand.
\noindent
For the transverse basis, we content ourselves with considering only those components that
are non-vanishing perturbatively in the massless limit:
\begin{align}
    T_2^\mu(p,k) &= \left[ p^\mu \left(k\cdot q\right) - k^\mu \left(p\cdot q\right) \right] \,
                     \left(\kslash + \pslash\right) \,, \nonumber\\
    T_3^\mu(p,k) &= q^2 \gamma^\mu - q^\mu \qslash\,,  \nonumber\\
    T_6^\mu(p,k) &= \gamma^\mu \left(p^2 - k^2\right) + \left(p+k\right)^\mu \qslash\,,  \nonumber\\
    T_8^\mu(p,k) &= -\gamma^\mu k^\nu p^\lambda \sigma_{\nu\lambda} + k^\mu \pslash - p^\mu \kslash\,.
\label{eq:Ts}
\end{align}

The remainder of this paper is devoted to considering solutions of the DSE resulting from different choices of vertex arising from
various forms for these transverse components $\tau_i(k,p)$.
The chief ingredients will be the matching to perturbation theory in the appropriate limit and the preservation of
multiplicative renormalizability by the truncation scheme.

\section{\label{sec:MR} Multiplicative Renormalizability and Choice of  Vertex}

Early on in the course of investigating possible vertex truncations in the Dyson--Schwinger equations, it was noted that
multiplicative renormalizability was not guaranteed to be preserved~\cite{collins, zuber, Curtis:1990zs}.
Indeed, whilst not surprising in the case of a
bare vertex this problem remains even on adoption of the Ball-Chiu form for the fermion-photon vertex. It
became apparent that to satisfy this necessary property of the equations, one must include a transverse part in the
vertex. However, the WGTI does not furnish us with any information here so we must find other means to constrain these
transverse components.

\subsubsection{\label{sec:CP vertex} Curtis--Pennington Vertex}

One approach to obtain non-perturbative constraints on the transverse part of the vertex is to use perturbation
theory, and demand that the leading and sub-leading logarithms contained within our wave-function renormalization and
mass functions re-sum correctly. This was the guiding principle taken by Curtis and Pennington \cite{Curtis:1990zs} that led to the
following form for the transverse part of the vertex in quenched QED: all $\tau_{i} = 0$, except
\begin{align}
   ( \tau_6)^M = -\frac{\lambda_2^M(p^2,k^2)\,(k^2 + p^2)\,(k^2-p^2)}{\left(k^2-p^2\right)^2 + \left[M^2(k^2) + M^2(p^2)\right]^2}\,.
   \label{eq:CP}
\end{align}

Numerical studies employing the Curtis--Pennington (CP) vertex showed that not only were the equations
manifestly multiplicatively renormalizable for large values of the coupling, but also that they exhibited a much
milder violation of gauge invariance \cite{Curtis:1993py} in critical studies. The success of this vertex and  the lack of any
further developments over the years means that it has been employed in many studies, from QED in three dimensions \cite{Burden:1993gy, Fischer:2004}
to studies of QCD \cite{Fischer:2003, Alkofer:2004}.

\subsubsection{\label{sec:KP vertex} K{\i}z{\i}lers{\" u}-Pennington Vertex}

The K{\i}z{\i}lers{\"u}-Pennington (KP) vertex \cite{Kizilersu:2009kg} is an unquenched vertex in the sense that it has the right structure to satisfy both 
fermion and photon DSEs. Moreover it respects gauge invariance, multiplicative renormalizability, agrees with perturbation theory in the 
weak coupling limit and is free of kinematic singularities.
The appearance of logarithms of the fermion finite renormalization function $A(p^2)$ here are a result of the requirement that, for a perturbative
expansion in the coupling, the coefficients of the leading logarithms exhibit the correct dependence on one another.
The KP vertex has the following construction for the four unknown transverse form factors:
\begin{align}
\tau_2^E =& - \frac{4}{3}\,\frac{1}{(k^4-p^4)}\,\left(A_k\,-\,A_p\right)\nonumber\\
                                   &- \frac{1}{3} \,\frac{1}{(k^2+p^2)^2}\,  \left(A_k + A_p \right)\,\ln \left[\left(\frac{A_k\,A_p}{A_q^2}\right)\right]\,,  \nonumber\\
\tau_3^E=&- \frac{5}{12}\, \frac{1}{(k^2-p^2)}    \; \left(A_k\,-\,A_p\right) \nonumber\\
                                   &-\frac{1}{6}\,\frac{1}{(k^2+p^2)}\;  \left(A_k + A_p \right)\,   \ln \left[\left(\frac{A_k\,A_p}{A_q^2}\right)\right] \,, \nonumber\\
\tau_6^E = & \frac{1}{4} \;\frac{1}{(k^2+p^2)}\,   \left(A_k\,-\,A_p\right)\,, \nonumber\\
\tau_8^E = &0\,,
\label{eq:finaltaus3}
\end{align}
where `E' denotes Euclidean space, and the momentum arguments of the $\tau_i^E(p^2,k^2,q^2)$ have been suppressed. $A_k$ is a convenient shorthand for $A(k^2)$. Later we adopt a similar shorthand for $\alpha(\mu^2)$ of $\alpha_\mu$.

\section{\label{sec:QUENCHED QED}Quenched QED$_4$ And Dynamical Mass Generation}

If we ignore the contribution of fermion loops to the photon propagator, in effect quenching the theory, the
coupling no longer depends on the photon momentum.
 A consequence of this truncation is that we do not need to consider the renormalization
of the theory, since the only scale that enters the problem is the numerical cut-off $\Lambda$. It then makes sense to work in
terms of dimensionless quantities ${\hat{p}^2}=p^2/\Lambda^2$ and $M({\hat{p}})/\Lambda$. Despite this displeasing scale dependence, 
the same constraints of multiplicative renormalizability may still be formally applied.

In this section, we consider three different choices of fermion-photon vertex; the bare vertex, the Curtis--Pennington vertex
 and the K{\i}z{\i}lers{\" u}-Pennington vertex. The last two vertices include the Ball-Chiu construction as the longitudinal vertex.
To study the impact of the truncation scheme (i.e vertex Ansatz and cutoff regulator) on the breaking of gauge invariance, we
perform our calculations in three representative gauges, $\xi =0, 1, 3$, and examine the behaviour of the critical coupling.
 Only Eqs.~(\ref{eq:ferwavefunc}, \ref{eq:massfunc}) of the DSEs need to be solved, since $G=1, N_F=0$ in the quenched approximation.

\subsection{\label{sec:Quenched QED and Dynamical Mass Generation} Previous Studies}
Dynamical mass generation in strongly coupled QED in 3- and 4- dimensions has been historically of great interest. However the majority of the previous studies in the literature have been limited to the quenched approximation of the theory.  The advantage of this approximation is that one can study the dynamical fermion mass generation analytically as well as numerically with some vertex approximations.  Undeniably  the results and guidance provided by these studies have been very valuable in developing our understanding of the phenomena of dynamical mass generation and for advancing the investigation of the robustness, performance and reliability of the numerical treatment~\cite{Schreiber:1998ht, Bloch:1995dd, Bloch:1994if,Williams:2007zzh}.

We now discuss some of the outcomes of these studies \cite{Fomin:1976af,Miransky:1984ef,Miransky:1986xp,Miransky:1986ib,Miransky:1979ks,Fukuda:1976zb,Fomin:1984tv,Lombardo:1994vz,Atkinson:1990bg,Atkinson:1986aw,Atkinson:1993mz,Curtis:1993py,Curtis:1992jg,Kizilersu:2000qd,Kizilersu:2001,Bloch:1994if,Bloch:1995dd,Williams:2007zzh,Bashir:1994az,Bashir:2011dp,Gusynin:1998se,Hawes:1991,Hawes:1994ce,Hawes:1996mw,Williams:1995rd,Reenders:1999fz,Fomin:1984fd} in 4-dimensional quenched QED. 
The quenched DSE investigation of dynamical mass generation  suggests that QED$_4$ undergoes a phase transition at a critical coupling
$\alpha_c$, when the interaction is strong enough. In the absence of a bare mass in the Lagrangian, the fermions in the theory are massless for all couplings less than this critical value $(\alpha < \alpha_c)$, while they acquire mass for couplings greater than the critical one $(\alpha > \alpha_c)$. The value of this critical coupling strongly depends on the truncation of the system~\cite{Oliensis:1990sg,Miransky:1984ef,Miransky:1986xp,Fukuda:1976zb,Fomin:1984tv,Lombardo:1994vz,Atkinson:1990bg,Atkinson:1986aw,Atkinson:1993mz,Curtis:1993py,Curtis:1992jg,Kizilersu:2000qd,Kizilersu:2001,Bloch:1994if,Bloch:1995dd,Williams:2007zzh,Bashir:2011dp,Gusynin:1998se,Hawes:1991,Hawes:1994ce,Hawes:1996mw} 
and which fermion-photon vertex is used. Furthermore it should be gauge independent~\cite{Curtis:1990zr,Atkinson:1993mz,Bashir:1994az,Bloch:1995dd,Williams:2007zzh,Bashir:2011dp} since it is (at least, in principle) a physical quantity. 

In the very close vicinity of the critical point the dynamically generated mass admits  a power-law behaviour~\cite{Bashir:1995qr} in Euclidean space, $M(p^2) = (p^2)^{-s}$, since the SD system in quenched QED$_4$ is scale invariant.  The exponent, $s$,  which determines the asymptotic behaviour of the mass function is related to the anomalous dimension of the $\overline{\Psi} \Psi$ operator.  Although this operator is not relevant in the perturbative region of QED$_4$ where it violates the renormalizability of the theory, it becomes large and renormalizable in the nonperturbative region; hence it becomes a relevant operator in non-perturbative DSE studies~\cite{Leung:1989hw,Miransky:1989nu,Gusynin:1997cw,Reenders:1999fz}.
This implies a four-fermion interaction with associated coupling parameter needs to be added to the theory.
Such operators are automatically included in 
lattice calculations which makes it difficult to compare lattice studies directly with the SD calculations that do not include such an interaction term. Nevertheless lattice studies qualitatively support the DSE findings of quenched QED$_4$ by observing that the theory goes through a phase transition breaking chiral symmetry with associated dynamically generated mass
~\cite{Kondo:1989my,Kondo:1988qd,Lombardo:1994vz,Kogut:1993ig,Kogut:1994dk,Kogut:1988ia,Kocic:1990fq,Kocic:1992vt}.
All of these studies agree that the dynamically generated mass obeys a mean field scaling law.

\begin{table}
 \caption{Critical couplings from previous studies for three different vertex Ans\"{a}tze for $\xi=0,1,3$ in quenched QED. The $^*$ vertex superscript indicates those solutions derived by applying the WGTI. (\lq NA' indicates \lq not available'.)}
 \label{table:critval_quenched_hist}
\begin{ruledtabular}
\begin{tabular}{l| c | c | c | c}
    $ \,\,\xi\,\,$                    &  0                                    &  1                                       &   3                                         &     Vertex      \\
\hline\hline
\,\,Ref.~\cite{Miransky:1984ef}\,\,    &  $\pi/3$                           & NA                                          &NA                                                  &    Bare  \\
\hline
\,\,Ref.~\cite{Curtis:1993py}\,\,     &  1.003\,*$\pi/3$                      &NA                                                &NA                                                            &    Bare  \\
\hline
\,\,Ref.~\cite{Bloch:1995dd}\,\,               &  1.047                             &1.690                                         &2.040                                               &    Bare  \\
\hline

\,\,Ref.~\cite{Curtis:1993py}\,\,    &  0.9344                          &0.9240                                        &0.9218                                              &    CP   \\
\hline
\,\,Ref.~\cite{Atkinson:1993mz}\,\, & 0.933667                       & 0.923439                            &0.921272                                      &    CP   \\
\hline
\,\,Ref.~\cite{Bloch:1995dd}\,\,               &  0.933667                       &0.890712                                   &0.832927                                           &    CP$^*$   \\
\hline
\,\,Ref.~\cite{Bashir:2011dp}\,\, &  0.934                            & NA                                                       &  NA                                                            &    BBCR Ansatz  \\
\end{tabular}
\end{ruledtabular}
\end{table}

Table \ref{table:critval_quenched_hist} shows critical couplings collected from various quenched QED$_4$ Dyson--Schwinger studies employing various vertices and using cutoff regularization. 
A few comments are in order. Miransky \cite{Miransky:1984ef} demonstrated analytically that dynamical mass generation occurs in Rainbow
QED (quenched QED with a bare vertex) in Landau gauge with a critical coupling of $\pi/3$. This was confirmed numerically by the other Rainbow studies cited. However, the use of the bare vertex makes the critical coupling highly gauge dependent, not least since it does not respect the WGTI.
The development of the CP vertex led to the numerical study of \cite{Curtis:1993py} and the analytic study of \cite{Atkinson:1993mz},
in substantial agreement, and exhibiting a reduced gauge dependence of the critical coupling. However, it emerged that using a cutoff
regulator potentially violates the translation invariance of the theory, and leads to an ambiguity in the fermion DSE equation
(except in Landau gauge), depending on whether or not the WGTI was applied in its derivation \cite{Curtis:1993py,Atkinson:1993mz},
which was resolved using dimensional regularization \cite{Schreiber:1998ht} in favour of the former scenario.
This accounts for the differing results for the CP vertex for $\xi \neq 0$ in Table \ref{table:critval_quenched_hist}.
Curiously, solutions for the CP vertex with the WGTI identity (correctly) applied exhibit greater gauge dependence
than those where it is not applied.

In summary, quenched QED using the bare and CP vertex is now well understood. However, inspection of
Table \ref{table:critval_quenched_hist} reveals that the desired gauge independence of the critical coupling has only been
partially realized, although the CP vertex represents a considerable improvement over the bare vertex~\cite{Curtis:1993py}.

\begin{figure*}[!t]
\subfigure[Bare Vertex]{
\includegraphics[width=0.32\textwidth]{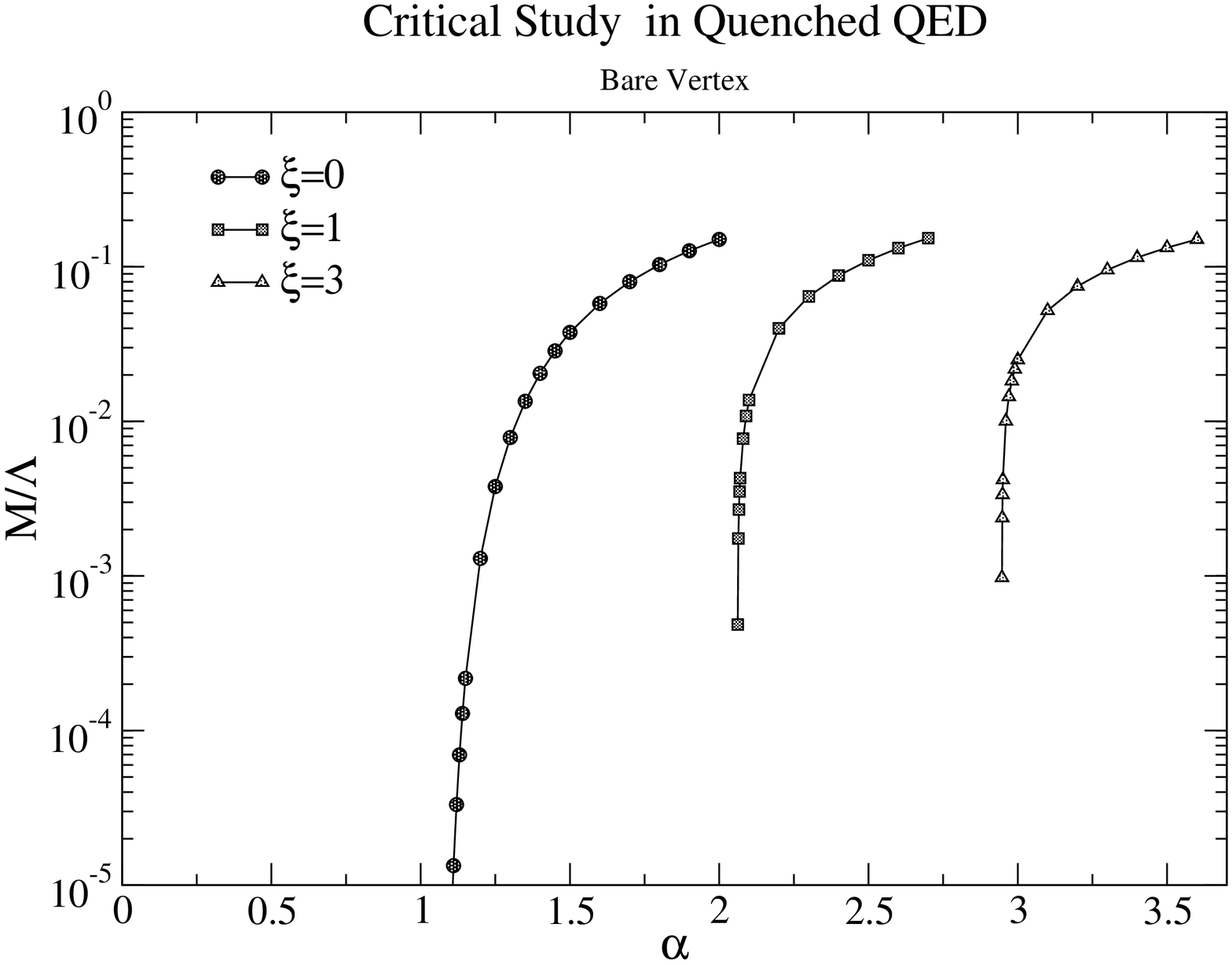}}
\subfigure[Curtis--Pennington Vertex]{
\includegraphics[width=0.32\textwidth]{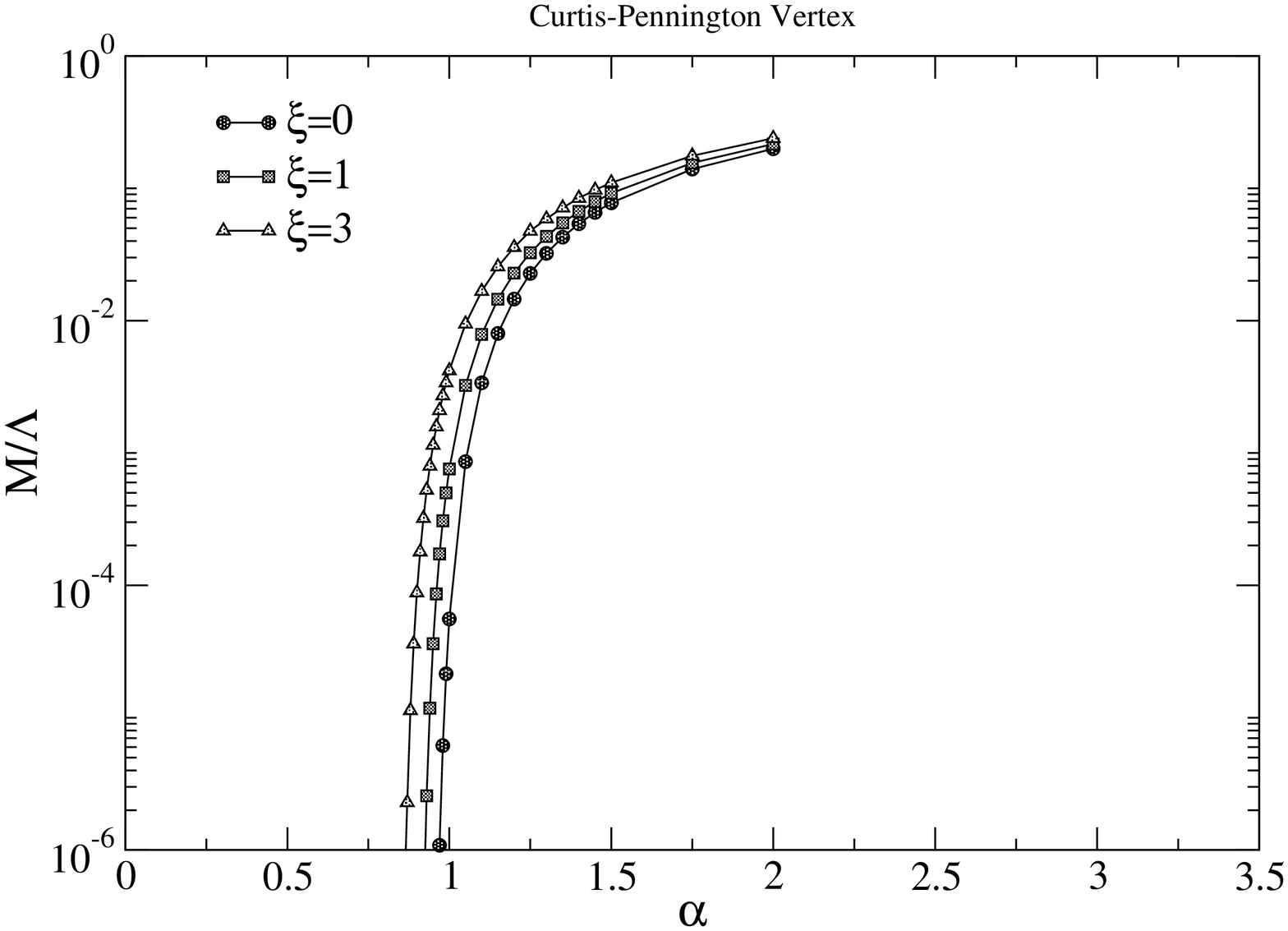}}
\subfigure[Kizilersu--Pennington Vertex]{
\includegraphics[width=0.32\textwidth]{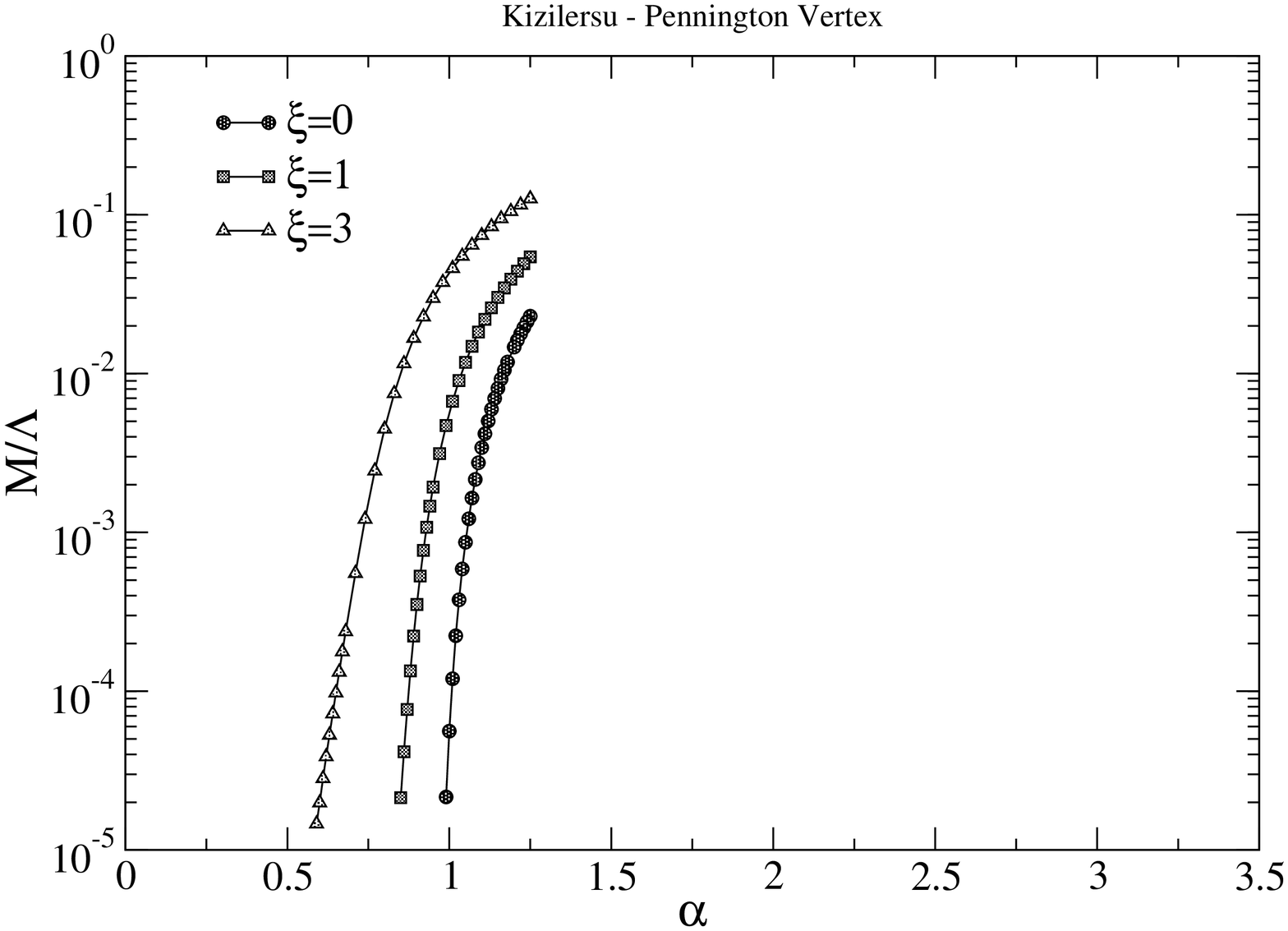}}
\caption{Dynamically generated Euclidean mass versus coupling for $\xi=0, 1, 3$ using the bare, Curtis--Pennington and \kizilersu--Pennington vertices respectively, without the WGTI.}
\label{fig:quenched_mass_gen}
\end{figure*}

\subsection{\label{sec:Quenched QED Numerical Results} Numerical Results for the Quenched KP Vertex}

Here, we study the critical behaviour of quenched QED$_4$ using the K{\i}z{\i}lers{\" u}-Pennington
vertex~\cite{ Kizilersu:2009kg,Williams:2007zzh} in various gauges, comparing it with the bare and CP vertices.
Only Eqs.~(\ref{eq:ferwavefunc}, \ref{eq:massfunc}) of the DSEs need to be solved, since $G=1, N_F=0$ in the quenched approximation.
We show our results in Fig.~\ref{fig:quenched_mass_gen} by plotting the dynamically generated Euclidean mass versus the coupling for $\xi =0,1,3$. We expect the Euclidean mass near criticality to be only approximately gauge invariant since it is different than the physical mass which should be exactly gauge-independent : however the location of the 
critical coupling should be gauge invariant. Both solutions with and without application of the WGTI were run: Fig.~\ref{fig:quenched_mass_gen} shows those without the WGTI.  In these studies the momentum cut-off is $\Lambda^2 = 10^{10}$. For comparison purposes, we repeat these calculations using the bare and CP vertices.
Since the mass-function exhibits an infinite order phase transition, the measure of dynamical mass generation obeys
the Miransky scaling law \cite{Fomin:1984tv}. When the coupling is greater but very close to its critical value, we have 
\begin{align}
\frac{\Lambda}{M_E} = \exp\left(  \frac{A}{\sqrt{\frac{\alpha}{\alpha_C} - 1}} -B \right)\,,
\label{eq:Miransky}
\end{align}
where $A, B$ and $\alpha_c$ can be determined through a least-squares fit.
We summarize these critical couplings in Table \ref{tab:quenchedvertexcomp} which show the results for solutions with the WGTI applied.

\begin{figure}[htpb]
 \includegraphics[width=0.80\columnwidth,angle=-90]{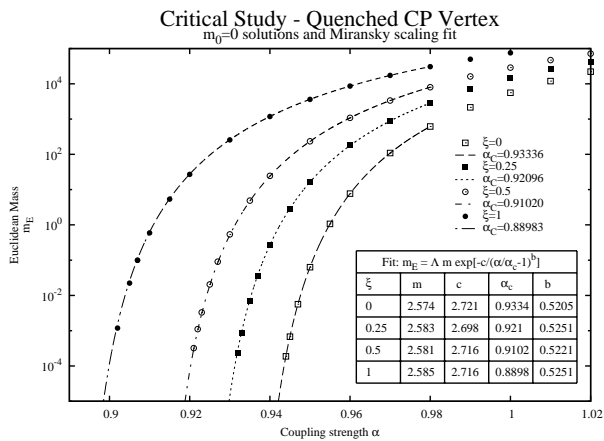}
 \caption{Dynamical mass versus coupling for $m_0=0$ solutions in quenched QED$_4$ for gauges $\xi=0, 0.25, 0.5, 1$ using the CP vertex with WGTI.}
 \label{fig:cp-massgen-cvg-fit-1}
 \end{figure}

\begin{table}[!b]
  \caption{Critical couplings for three different vertex Ans\"{a}tze for $\xi=0,1,3$ in quenched QED, with the WGTI applied.
The Miransky scaling law, Eq.~(\ref{eq:Miransky}), is used to extract $\alpha_c$.
These results were generated by the Durham group~\cite{Williams:2007zzh}.}
  \label{tab:quenchedvertexcomp}
\begin{ruledtabular}
  \begin{tabular}{l| c | c | c | c}
$ \,\,\xi\,\,$                    &  0         &  1          &   3         &     Vertex       \\
\hline
\hline
$\,\,\alpha_c\,\,$               &  1.0472 & 1.690  &  2.040 &    Bare vertex \\
\hline
$\,\,\alpha_c\,\,$               &  0.9339 & 0.8909  &  0.8329 &    Curtis--Pennington  \\
\hline
$\,\,\alpha_c\,\,$               &  0.9351 & 0.7222  &   $\simeq$ 0.5       & \kizilersu--Pennington  \\
\end{tabular}
\end{ruledtabular}
\end{table}
Figure \ref{fig:cp-massgen-cvg-fit-1} shows the dynamically generated mass versus coupling for the CP vertex for $\xi=0, 0.25, 0.5, 1$, with fit parameters $m$, $c$, $\alpha_{c}$ and $b$ in the formula
 \begin{align}
 	\frac{m_E}{\Lambda}=m \exp\left[-\frac{c}{(\alpha/\alpha_c-1)^b}\right] \,.
 \end{align}

It is  apparent that both the CP and KP vertices outperform the
bare vertex, though the CP vertex has the smallest degree of gauge variance of the three models considered.
Note that as the gauge parameter increases, the value of the critical coupling decreases for the CP and KP vertices,
and increases for the bare vertex.  

\section{\label{sec:Massless Unquenched QED}Massless Unquenched QED$_4$ }

Introducing fermion loops into the Dyson--Schwinger equation for the photon propagator leads to a running of the
coupling.  Renormalization then becomes mandatory, allowing us to trade the cut-off $\Lambda$ for some physical renormalization
point, $\mu$.

Before investigating the effects of dynamical mass generation
by examining critical behaviour, we investigate the strictly massless theory. Our primary purpose is
to quantify the gauge-dependence of the photon dressing function $G(p^2)$, which should ideally be independent of $\xi$.
Additionally, we wish to consider the effect of using a cutoff regulator, which potentially violates the translation invariance
of the theory. The photon self energy diagram, Fig.~\ref{fig:DSE}, treats both the fermion propagators in a symmetric way, but this
symmetry will potentially not be respected by the cutoff regulator~\cite{AK:thesis,Williams:2007zzh,Kizilersu:2013hea};
this is investigated numerically below by introducing a fermion loop variable parameter $\eta$ with $k -\eta\,q$ the upper loop momentum
in Fig.~\ref{fig:DSE} and $k+(1-\eta)q$ the lower. Two cases were investigated:
\begin{itemize}
\item $\eta = 1$ (the {\em asymmetric partition}) corresponds to loop momenta $k-q$ and $k$
\item $\eta = 1/2$ (the {\em symmetric partition}) corresponds to loop momenta $k-q/2$ and $k+q/2$
\end{itemize}

\begin{figure*}[ht]
\subfigure{
\includegraphics[width=0.9\columnwidth]{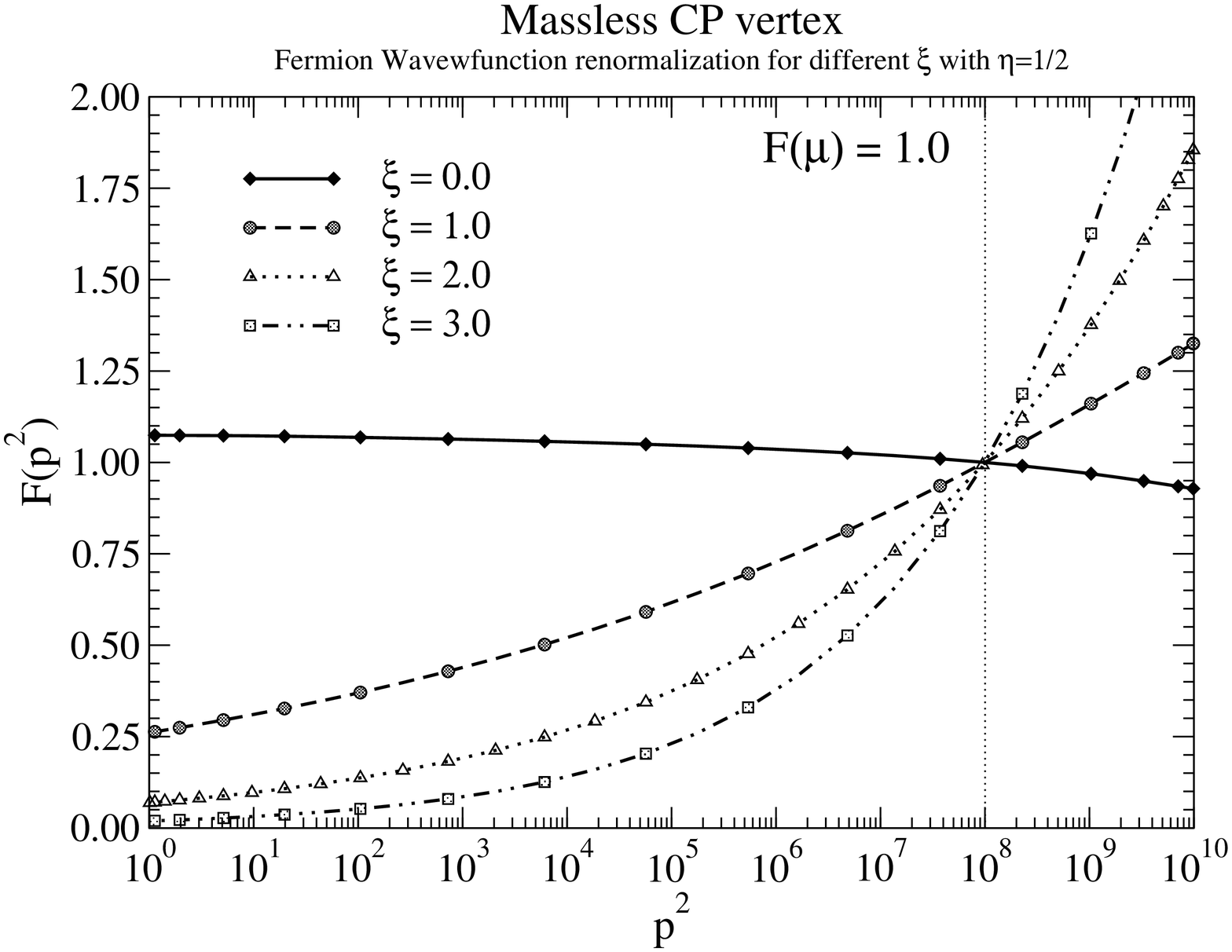}}\quad
\subfigure{
\includegraphics[width=0.9\columnwidth]{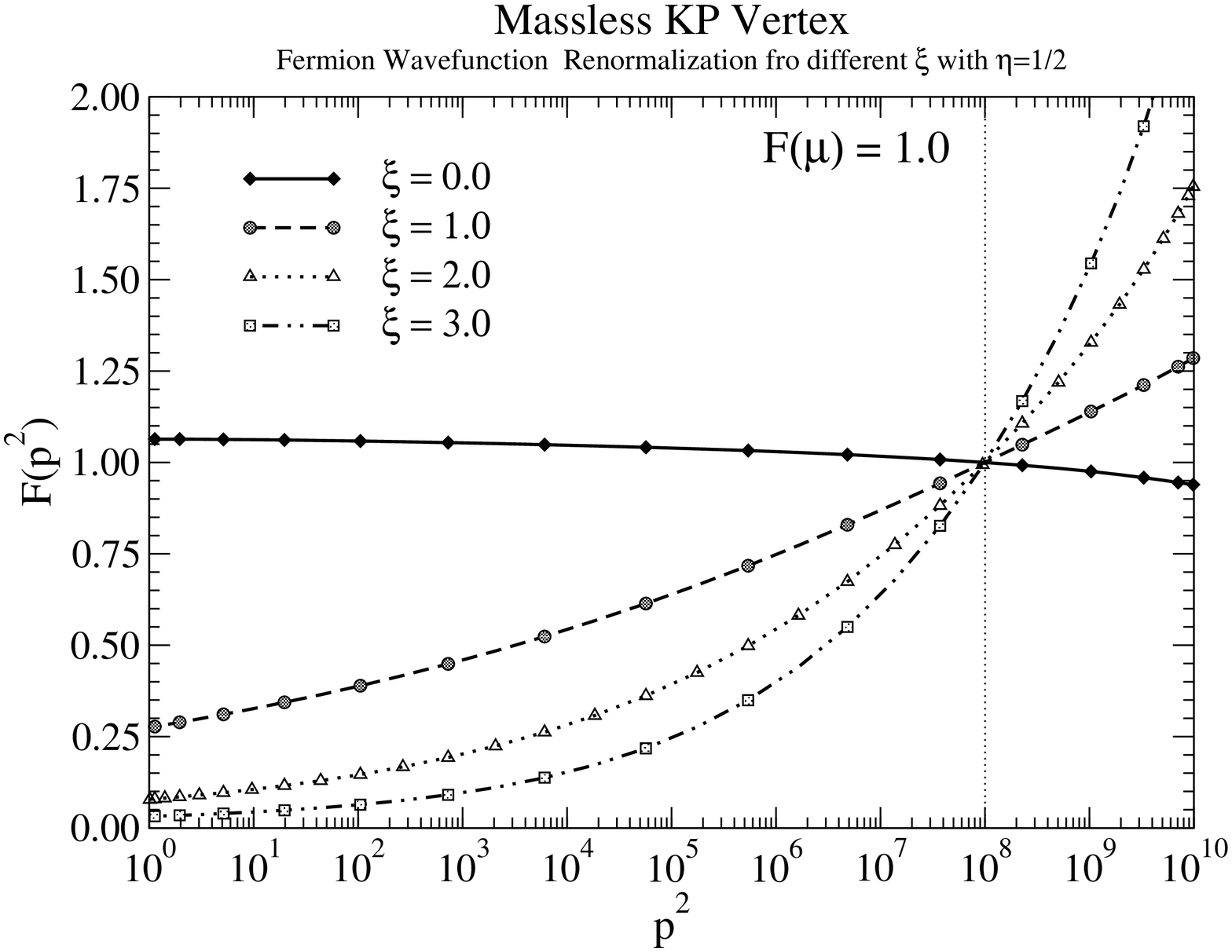}}\\
\subfigure{
\includegraphics[width=0.9\columnwidth]{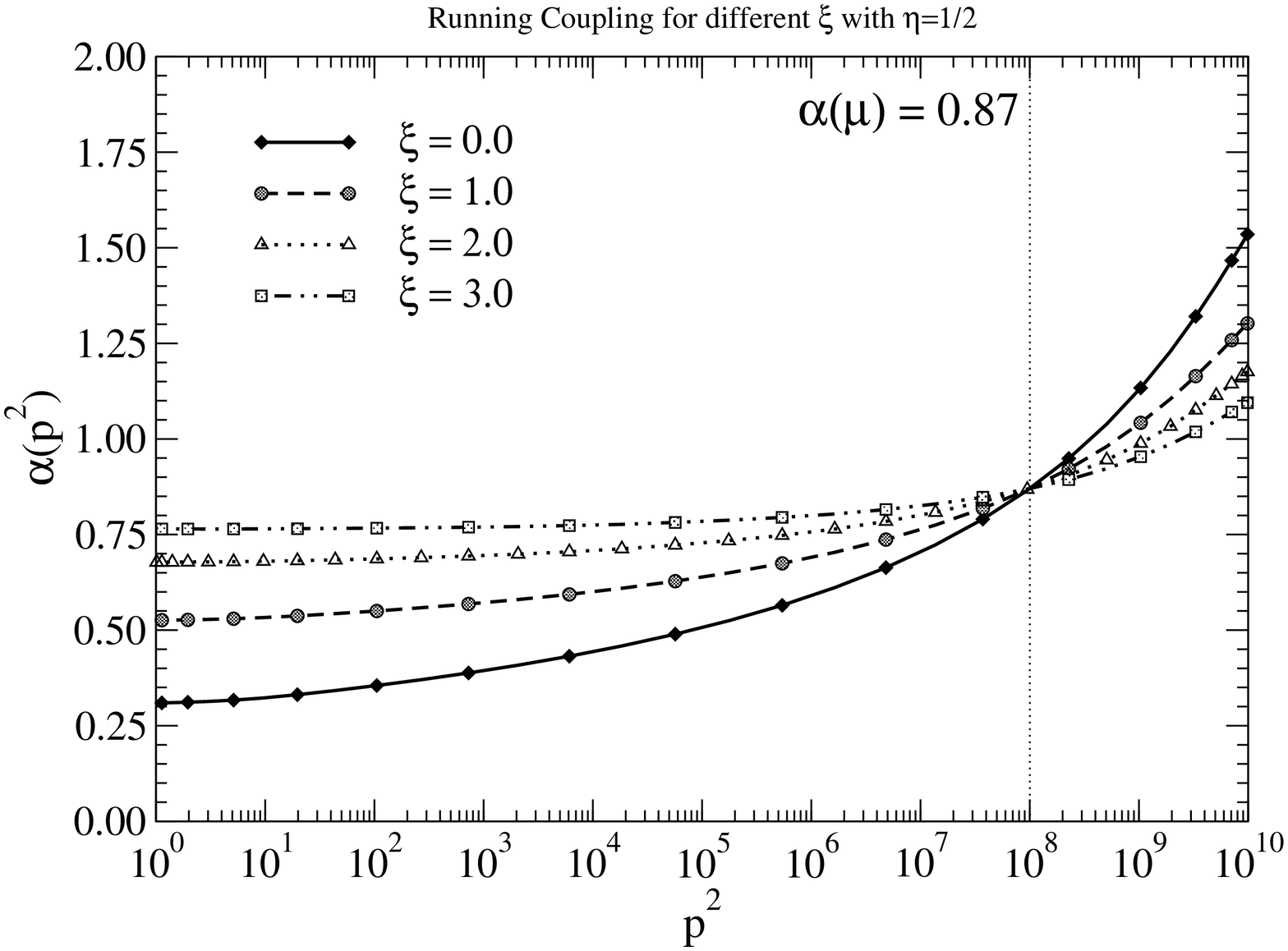}}\quad
\subfigure{
\includegraphics[width=0.9\columnwidth]{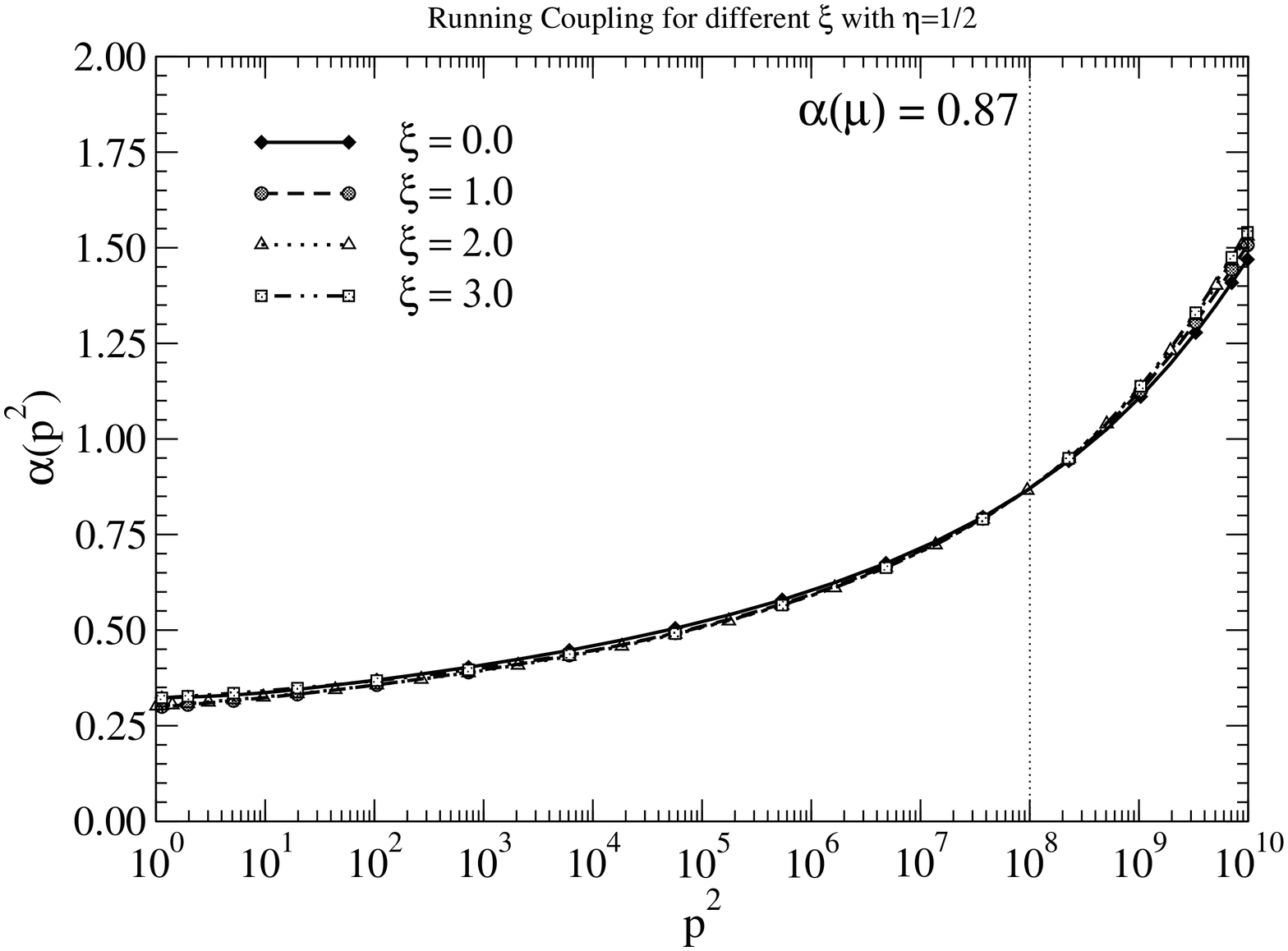}}
\caption{Fermion wave-function renormalization and running coupling for $\xi =0, 1, 2,3$ and $\eta=1/2$ (symmetric momenta partition) employing the Curtis--Pennington vertex (left) and 
K{\i}z{\i}lers{\"u}-Pennington (right).}
\label{fig:massless_F_alpha}
\end{figure*}

In Fig.~\ref{fig:massless_F_alpha}, we show the results of the fermion wave-function renormalization and the corresponding effective (running) coupling versus momentum-squared for four gauges, $\xi =0, 1, 2, 3$, and $\eta=1/2$ with the modified Curtis--Pennington vertex (left panels) and K{\i}z{\i}lers{\" u}-Pennington vertex (right panels). For both vertices the function $F$ is dependent upon the gauge, as it must be; the solutions are virtually identical. However one sees the photon dressing function is strongly gauge dependent for the CP vertex, motivating the development of a `better' vertex, the KP vertex, for which the improvement in the effective (running) coupling is instantly apparent.
The solutions for four different choices of the gauge parameter 
lie almost on top of one another. The requirement that the vertex Ansatz ensures the DSE should be multiplicatively renormalizable dramatically reduces the violation of gauge invariance.

\begin{figure*}[htpb]
\subfigure{
\includegraphics[width=0.8\columnwidth]{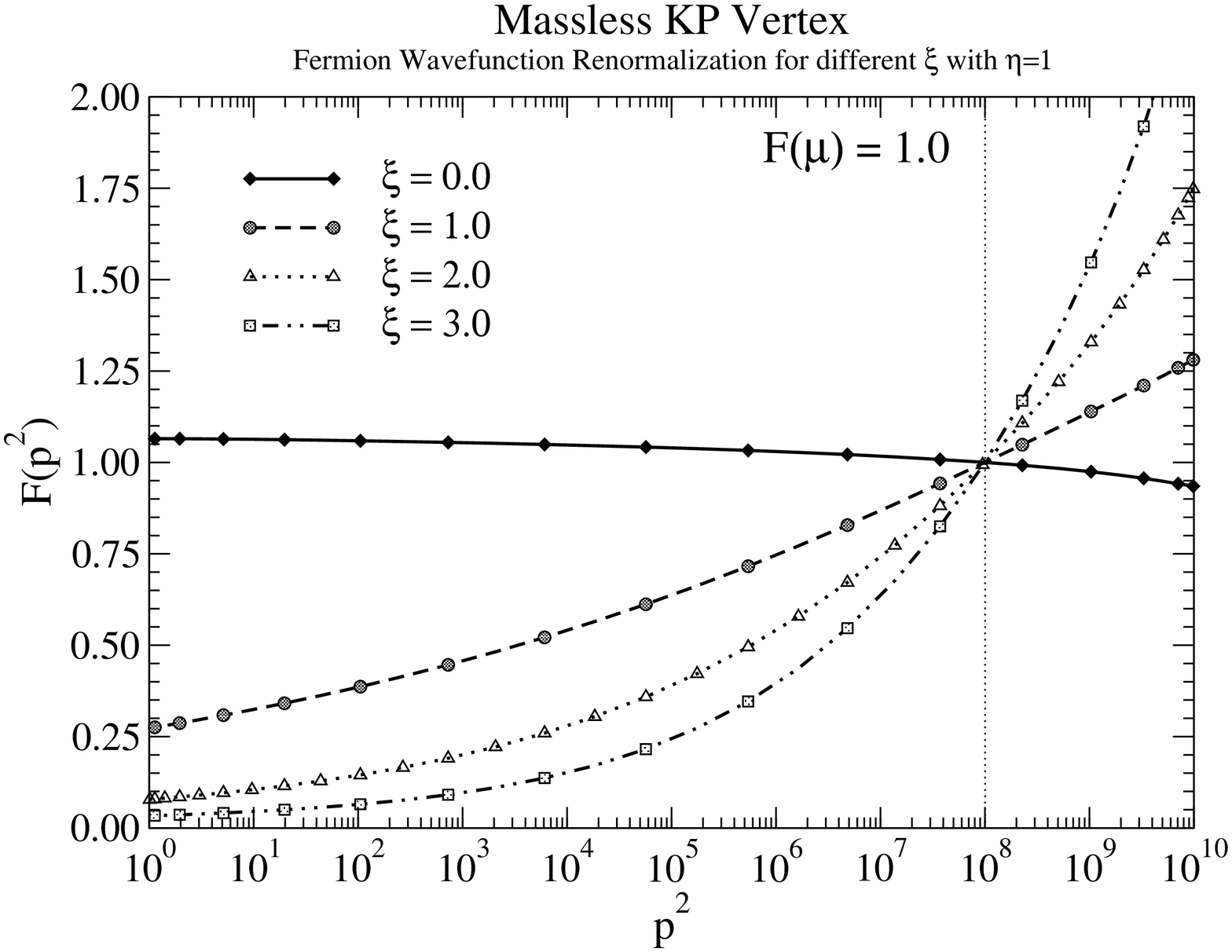}}
\subfigure{
\includegraphics[width=0.8\columnwidth]{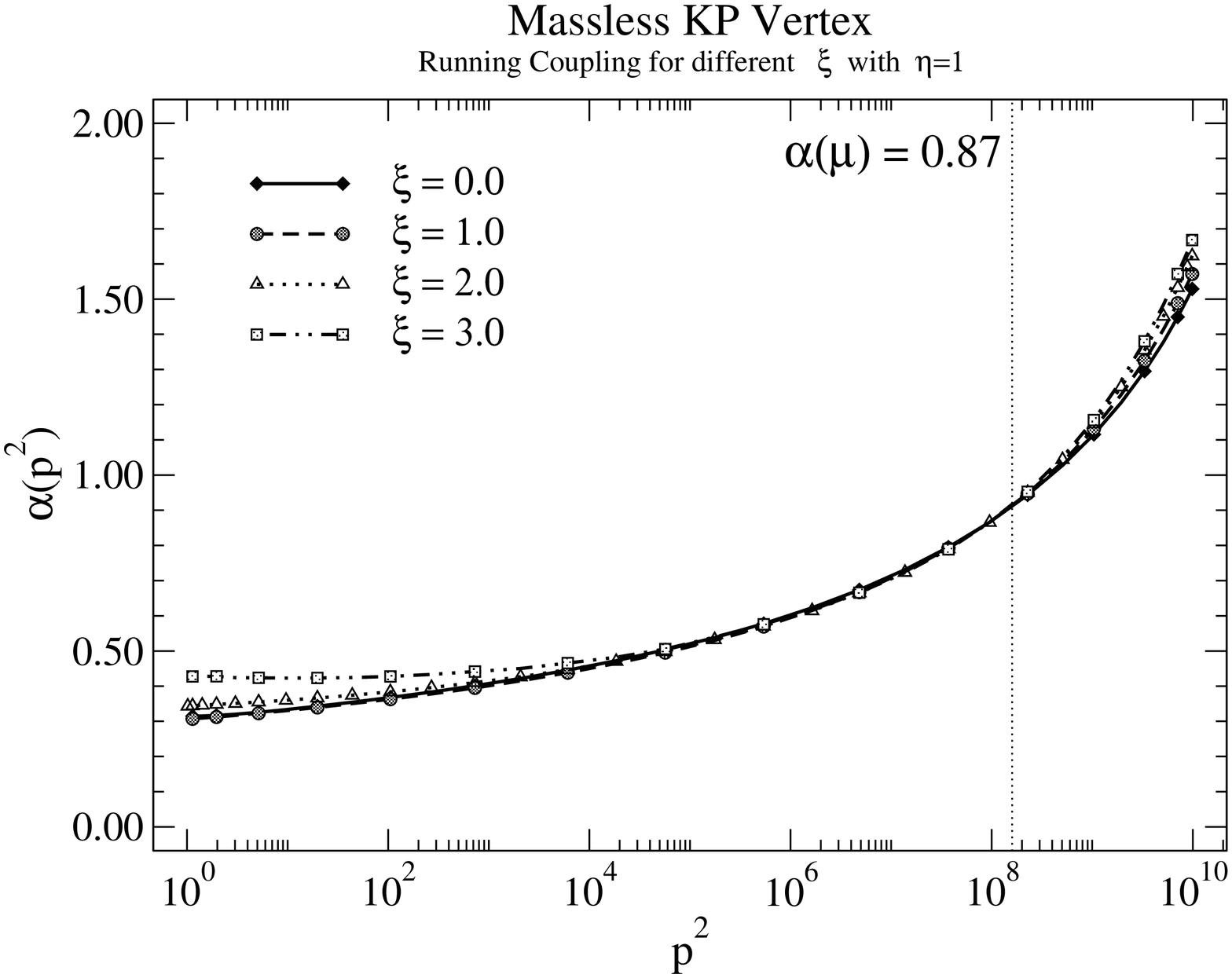}}
\caption{Fermion wave-function renormalization and running coupling for $\xi =0, 1, 2,3$ and $\eta=1$ (asymmetric momenta partition) employing the K{\i}z{\i}lers{\"u}-Pennington vertex.}
\label{fig:massless_F_alpha_asym}
\end{figure*}

Figure~\ref{fig:massless_F_alpha_asym} shows similar results as in Fig.~\ref{fig:massless_F_alpha}
for the KP vertex, but with the asymmetric momentum partition $\eta=1$.
The difference between the two momentum partition schemes is most apparent
in the running coupling in the infrared.  

\section{\label{sec:Massive Unquenched QED and Dynamical Mass Generation}Massive Unquenched QED$_4$ and Dynamical Mass Generation}

\subsection{Previous Studies}
While exploring fermion mass generation in the unquenched (full) theory is the ultimate goal,
historically most effort devoted to this subject using the DSE formalism has been conducted using various approximations which make the system tractable : for example, the bare vertex and/or the quenched approximation.
Fundamentally the aim of unquenching QED is to understand the effects of the fermion loops on the interaction,
namely the behaviour of the running coupling for a given physical system. There exists a large body of literature on dynamical fermion mass generation in unquenched QED$_4$~\cite{Gusynin:1989mc,Gusynin:1989ca,Kondo:1990ig,Kondo:1991ms,Kondo:1990ky,Kondo:1990ar,Kondo:1990st,Oliensis:1990sg,Ukita:1990uw,Bloch:1994if,Bloch:1995dd,Williams:2007zzh,Bashir:2011ij,Akram:2012jq,Kizilersu:2013hea} for various gauges $\xi$ and number of flavours $N_F$. For reasons explained above, these necessarily involve truncations in the vertex and/or propagators; the most popular truncations are to replace the full photon wave function renormalization with its 1-loop perturbative expression, avoiding angular integrations in the DSEs, and solving the DSEs by iteration for $F, M$ and $G$ with the bare vertex. Invariably, cutoff regularization is used. The results obtained are qualitatively similar to the quenched case; namely, a phase transition occurs at some critical value of the coupling whereby fermion masses are generated dynamically; and (more controversially) the equations obey scaling laws from which conclusions are drawn regarding the continuum limit of the theory.

Some of these critical value studies in the literature are tabulated in Table~\ref{tab:unquenchedvertexcomplit}.
Lattice studies also obtain dynamical mass generation and mass function scaling ~\cite{Aoki:1989bu,Dagotto:1990dd,Gockeler:1993ms,Gockeler:1993cu,Gockeler:1997dn,Gockeler:1990bc,Gockeler:1989wj,Kocic:1992vt,Kocic:1990fq,Kogut:1987cd}, but are not currently directly comparable because as explained above, they naturally contain a four-fermion interaction term.
 \begin{table}
 \caption{Critical couplings in the literature for different vertex Ans\"{a}tze for in Landau gauge in unquenched QED$_4$. This is an extension to Table 4.1 in Ref.~\cite{Bloch:1995dd}}
 \label{tab:unquenchedvertexcomplit}
 \begin{ruledtabular}
 \begin{tabular}{c| c | c }
 Ref.                                &  \,\,$\alpha_c (N_F=1)$\,\,         &  Vertex Model                         \\
\hline
\hline
\cite{Kondo:1990ig}        &   1.9997 $(\xi=0)$           &  Bare                                         \\
\hline
\cite{Gusynin:1989mc}     &   1.95 $(\xi=0)$              & Bare                                        \\
\hline
\cite{Kondo:1990st}       &  1.9989$(\xi=0)$              &  Bare                            \\
\hline
\cite{Kondo:1990st}       &  2.0728 $(\xi=0)$             &  Bare                             \\
\hline
\cite{Oliensis:1990sg}       & 1.9995 $(\xi=0)$             &  Bare                              \\
\hline
\cite{Rakow:1990jv}       &  2.25 $(\xi=0)$                 &  Bare                              \\
\hline
\cite{Atkinson:1990bg}    &  2.10028 $(\xi=0)$             &  Bare                            \\
\hline
\cite{Kondo:1991ms}      &  2.084 $(\xi=0) $            &  Bare                            \\
\hline
\cite{Ukita:1990uw}                     &    2.0944            &        NA                     \\
\hline
\cite{Kondo:1993aq}       &  1.9995             &   NA                               \\
\hline
\cite{Bloch:1995dd}    & 1.99953  $(\xi=0)$             &  Bare                          \\
\hline
\cite{Bloch:1995dd}    & 1.74102 @$\Lambda^2  (\xi=0) $            &  Bare                             \\
\hline
\cite{Bloch:1995dd}   & 1.63218 @$\Lambda^2  (\xi=0) $            &  Ball-Chiu                             \\
\hline
\cite{Bloch:1995dd}   & 1.61988 @$\Lambda^2  (\xi=0) $            &  modified CP                              \\
\hline
\cite{Bashir:2011dp}    &  2.27(Anal.), 2.4590(Num.) $(\xi=0)$                      &  BC+KP+A                             \\
\hline
\cite{Akram:2012jq}   &  0.9553 $(\xi=0)$                      &  BC+Ansatz                          \\
\end{tabular}

\end{ruledtabular}  
\end{table}
\subsection{Preliminaries}

In this investigation, we use the two vertex Ans\"{a}tze introduced in Section \ref{sec:MR}:
the Curtis--Pennington (CP) vertex and the K{\i}z{\i}lers{\" u}-Pennington (KP) vertex. 
However, we cannot use the former directly in massive studies since the transverse component of the CP vertex given in Eq.~(\ref{eq:CP})
leads to a quadratic divergence in the photon DSE, due to the DSEs probing different kinematical regions in the vertex
\cite{Kizilersu:2009kg}.
Instead, we use two models (we call them models since they are clearly unrealistic, but serve to illuminate possible paths forward)
where the CP vertex is used in the fermion DSE, and
\begin{itemize}
\item the Ball-Chiu construction without a transverse part is used in the photon DSE
\cite{Bloch:1995dd,Kizilersu:2013hea} (the {\em modified CP vertex})
\item the KP vertex is used in the photon DSE \cite{Williams:2007zzh} (the {\em modified KP vertex} or hybrid CP/KP vertex)
\end{itemize}

We also use cutoff regularization which should be recognized as an approximation,
although there are doubts whether QED makes sense without one.  We now proceed to
investigate the consequences of the choice of the vertex on dynamical mass generation, and the gauge invariance of the consequent critical coupling.
To accomplish this we introduce a sufficiently large
coupling into the theory such that our solutions exhibit dynamical mass generation in the absence of a bare mass, for a
selection of choices of the gauge parameter. The coupling strength is then decreased, in turn reducing the amount of
mass generation, until we cross into the phase where only a massless solution exists. We use a renormalized formalism previously
introduced in \cite{Kizilersu:2000qd,Kizilersu:2001}, but constrained to 
solutions with zero bare mass ($m_0=0$). In this case, the mass function Eq.~(\ref{eq:massfunc}) simplifies to
\begin{align}
	 M(k^2) = \frac {\overline{\Sigma}_s(k^2)}{1 - \overline{\Sigma}_d(k^2)}\,.
	 \label{eq:zerobaremass}
\end{align}
In this section, all solutions have $\mu^2=10^8$ and $\Lambda^{2} = 10^{10}$.

\begin{figure}[htpb]
\subfigure{
\includegraphics[width=0.8\columnwidth,angle=-90]{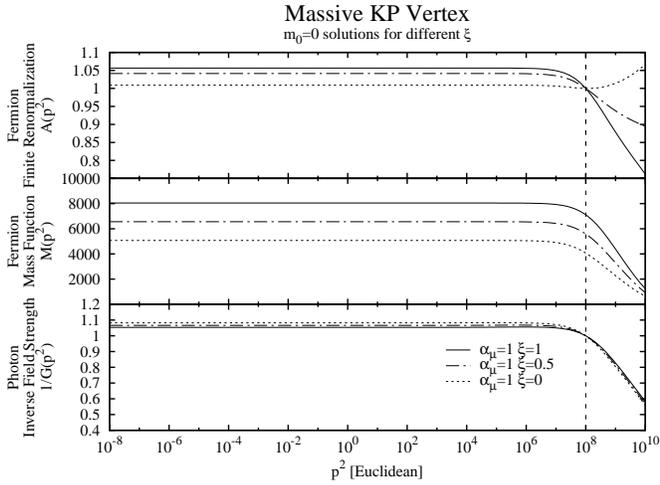}}
\subfigure{
\includegraphics[width=0.8\columnwidth,angle=-90]{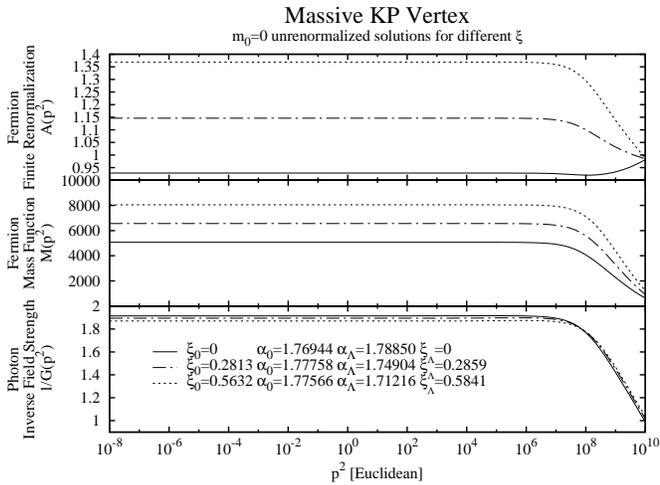}}
\caption{Renormalized $m_0=0$ solutions for $\alpha_{\mu}=1$, $\xi_\mu=0,0.5,1$ (top)
and corresponding unrenormalized solutions (bottom) for asymmetric momentum split $(\eta = 1)$ }
\label{fig:kp-chiral}
\end{figure}

 Typical fermion and photon propagator solutions for $m_{0} = 0$ and fixed $\alpha_{\mu}$ above criticality are presented 
 in Fig.~\ref{fig:kp-chiral} (top) for three different gauges ($\xi_{\mu} = 0, 0.5, 1$). In Eq.~(\ref{eq:zerobaremass}) 
 and Fig.~\ref{fig:kp-chiral} (top), $A(p^{2})$ and $G(p^{2})$ are renormalized solutions;
however their unrenormalized counterparts can easily be calculated, as they are proportional to the renormalized solutions.
Using Eqs.~(\ref{eq:renfermion}) and (\ref{eq:renphoton}) we get 
\begin{eqnarray}
	A_{0}(p^{2},\Lambda^2) &=& A(p^{2},\mu^2) \,\,/\,\, Z_{2}(\Lambda^2,\mu^2)	 \,,\\
        G_{0}(p^{2},\Lambda^2) &=& G(p^{2},\mu^2) \, Z_{3} (\Lambda^2,\mu^2)\,,
        \end{eqnarray}
the constants of proportionality are the fermion and photon renormalization constants, $Z_{2}$ and $Z_{3}$ respectively.
As the coupling constant transforms oppositely to $G$, Eq.~(\ref{eq:renalpha}), the unrenormalized coupling is
\begin{align}
	\alpha_{0} = \alpha \,\,/\,\, Z_{3} \,,
\end{align}
so that the {\em effective coupling} function
\begin{align}
	 \alpha_{\rm{eff}}(p^2) = \alpha(\mu^2) \, G(p^2,\mu^2) = \alpha_0(\Lambda^2) \, G_{0}(p^{2},\Lambda^2)\,,
\end{align}
is invariant. The mass function is similarly invariant, likewise $\alpha\,\xi$ is also an invariant quantity, so that $\xi$ transforms like $G$, and will also be shifted in the unrenormalized solutions (except for Landau gauge). 
The unrenormalized solutions corresponding to the top panel of Fig.~\ref{fig:kp-chiral} are presented in the bottom panel.
Also displayed are 
\begin{align}
	\alpha_\Lambda = \alpha_{\rm{eff}}(\Lambda^2) = \alpha \, G(\Lambda^2) \quad\mbox{and}\quad \xi_\Lambda = \xi/G(\Lambda^2)\;\;,
\end{align}
which are the parameters we would use for $\alpha$ and $\xi$ if we wished to obtain the same solution as in Fig.~\ref{fig:kp-chiral},
but renormalized at the cutoff (an example of a renormalization point transformation).
Note that $\alpha_{0} \approx \alpha_{\Lambda}$, and also $A(\Lambda^{2}) \approx 1$ and $G(\Lambda^{2}) \approx 1$: that is to say, the unrenormalized solutions are almost the same (but {\em not} identical) as solutions renormalized at the cutoff. 

\begin{figure}[!t]
\subfigure{
\includegraphics[width=0.8\columnwidth,angle=-90]{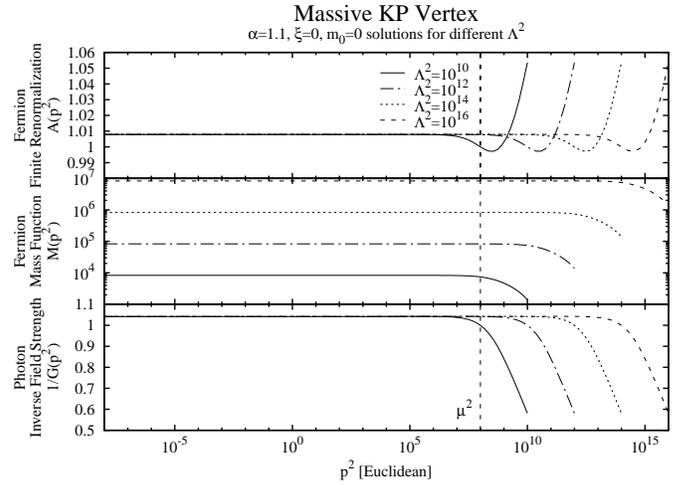}}
\subfigure{
\includegraphics[width=0.8\columnwidth,angle=-90]{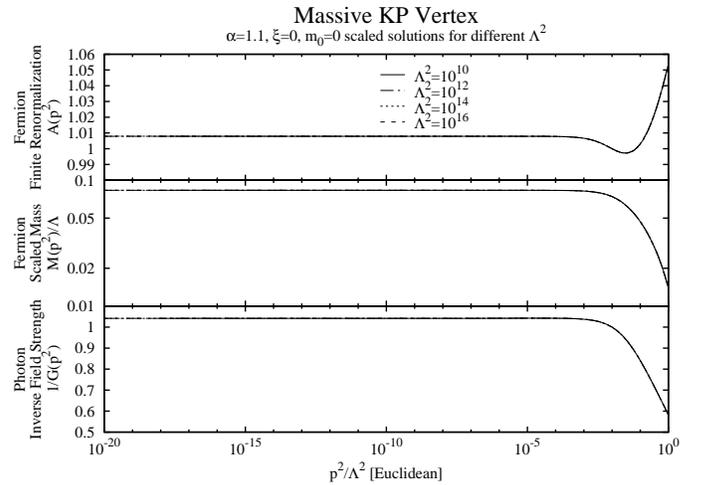}}
\caption{$m_0=0$ solutions for $\alpha_{\mu}=1.1$ and $\xi=0$, scaling $\Lambda^{2}$ and $\mu^2$ (top) and corresponding cutoff-relative solutions (bottom)}
\label{fig:kp-chiral2-allscale}
\end{figure}

We also note that the unrenormalized solutions will scale with the cutoff according to their mass dimension: that is to say, under
\begin{align}
	\Lambda \to \Lambda' \,,
\end{align}
$A_{0}$ and $G_{0}$ as functions of $p^2/\Lambda^2$ are unchanged (as is $\alpha_0$) and
\begin{align}
	M_{0} \to \sqrt{\Lambda'/\Lambda} \, M_{0}\;.
\end{align}
These relations can only be maintained using the renormalized solutions if the renormalization point is scaled with the cutoff. Figure~\ref{fig:kp-chiral2-allscale} illustrates the point~\cite{Curtis:1993py}. When all mass scales, here $\Lambda^2$ and $\mu^2$ (since $m_0 = 0$) are scaled simultaneously, the results are invariant when all momenta are plotted relative to the cutoff.

In the top  Fig.~\ref{fig:kp-chiral2-allscale} it is evident that, for zero bare mass solutions, the mass function scales with the cutoff. Furthermore, there is a general shift rightwards in momentum scale in all propagator functions, in proportion to the increase in cutoff.
It behaves the same as the quenched theory, as studied in \cite{Hawes:1996ig,Hawes:1996pe}. In these papers, it was argued that this was evidence that QED did not have a chiral limit in the usual sense. However, since the bare mass naturally varies with the cutoff, we should not expect bare mass solutions to be invariant against changes in the cutoff; rather the invariant solutions are the {\em scaled} solutions, as presented in the bottom Fig.~\ref{fig:kp-chiral2-allscale}.

These preliminary considerations show that it is viable to use a renormalized formalism to conduct a search for a critical coupling, below which the only solutions with $m_{0}= 0$ are massless, provided we quote the unrenormalized coupling or the coupling at the cutoff in our results. Comparison with renormalized solutions is only possible without conversion if they have the same cutoff {\em and} renormalization point, or at least the same $ratio$. We also note the (inconvenient) shift in $\xi$ for $\xi \neq 0$: this will vary with $\alpha$.

\subsection{Numerical Results}

\begin{figure*}[htp]
\subfigure{
\includegraphics[width=0.32\textwidth]{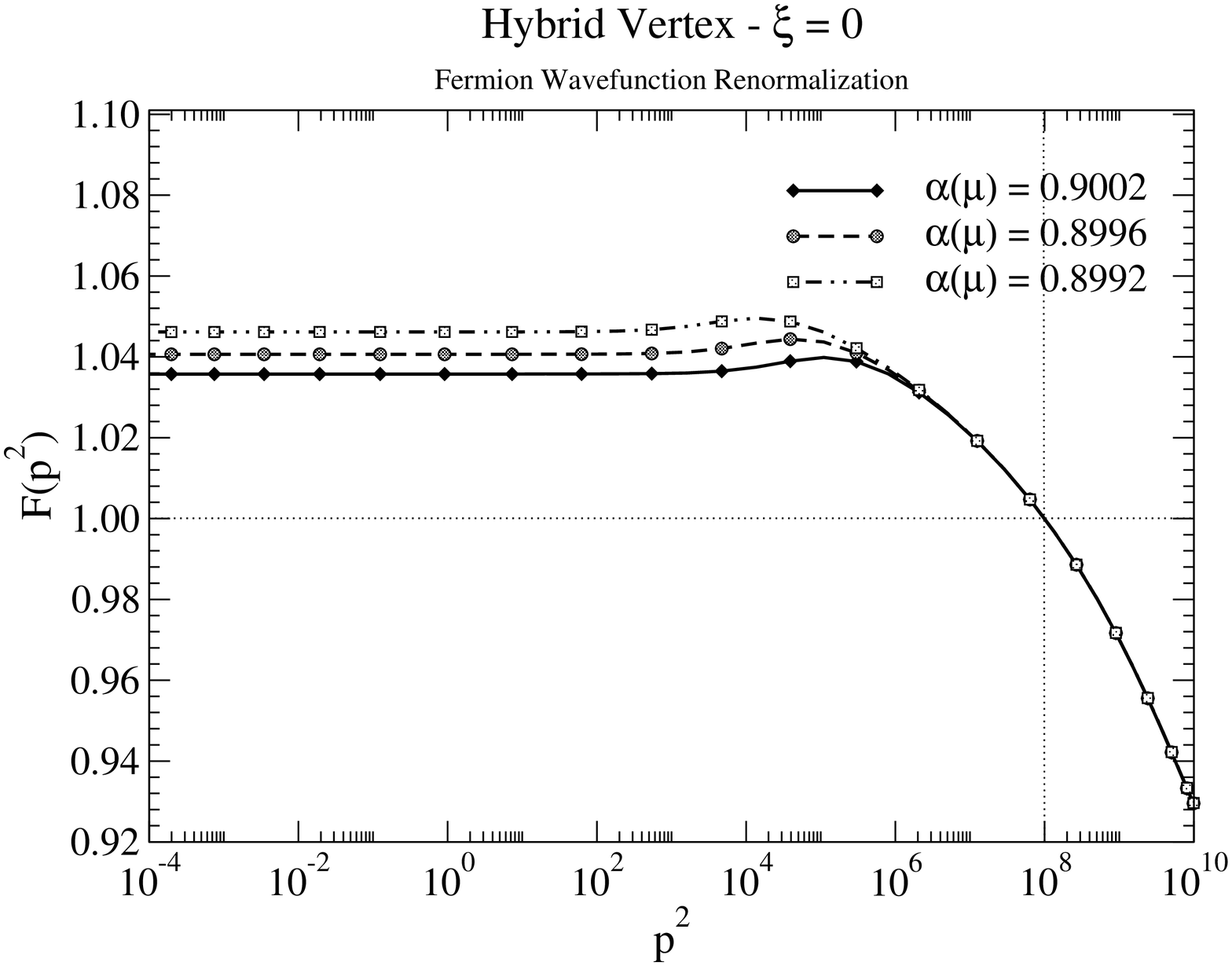}
\includegraphics[width=0.32\textwidth]{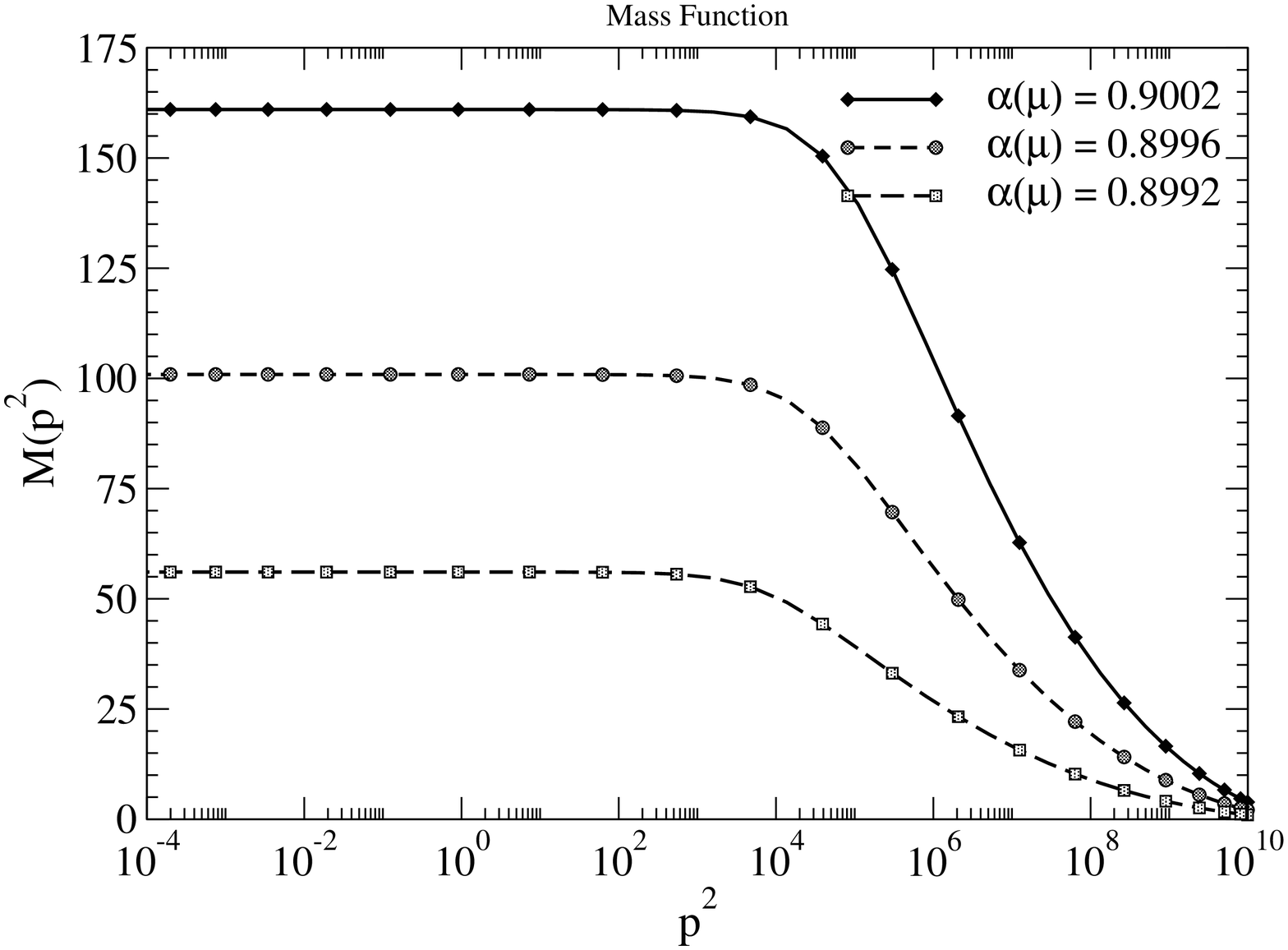}
\includegraphics[width=0.32\textwidth]{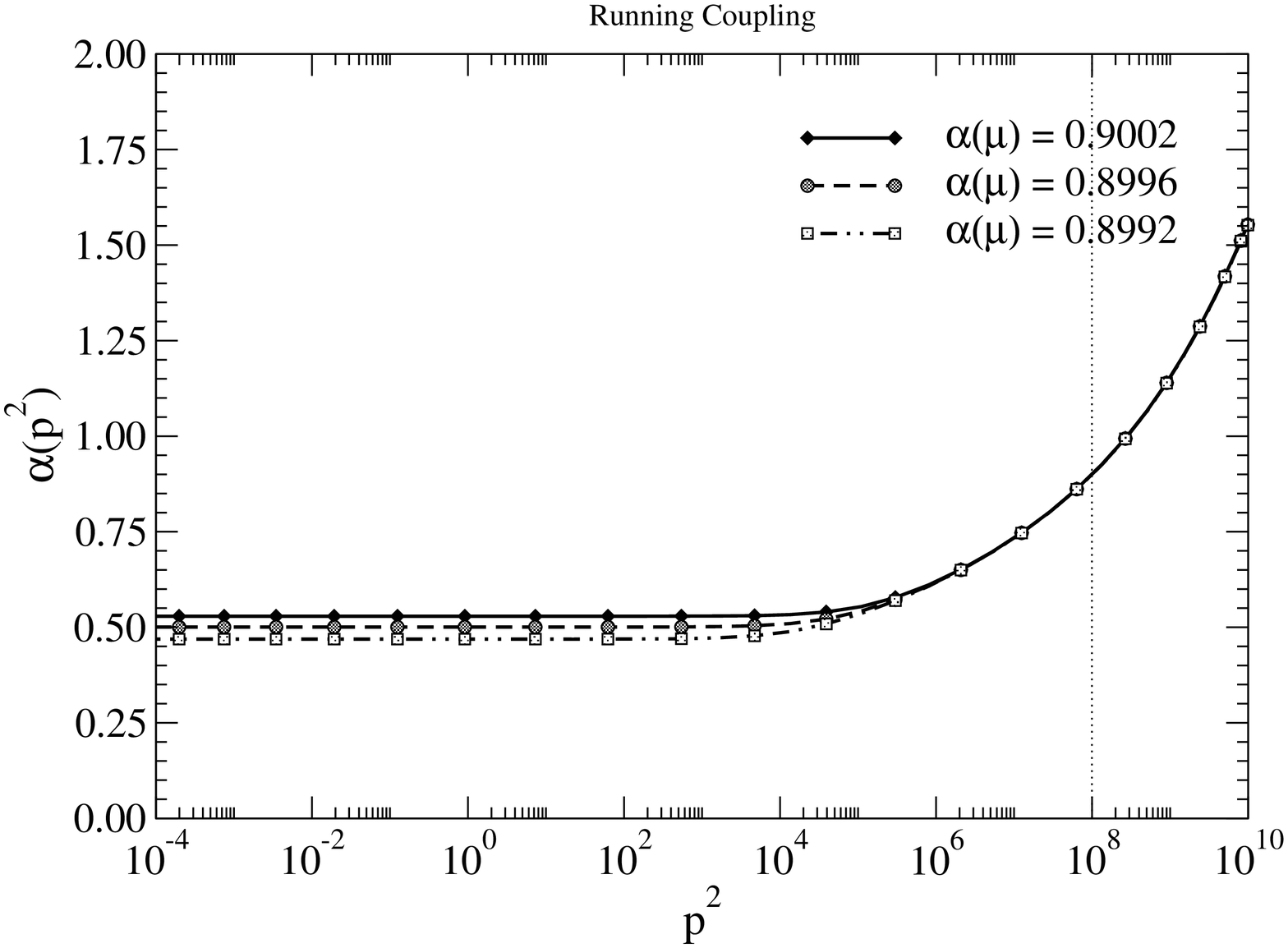}
}
\subfigure{
\includegraphics[width=0.32\textwidth]{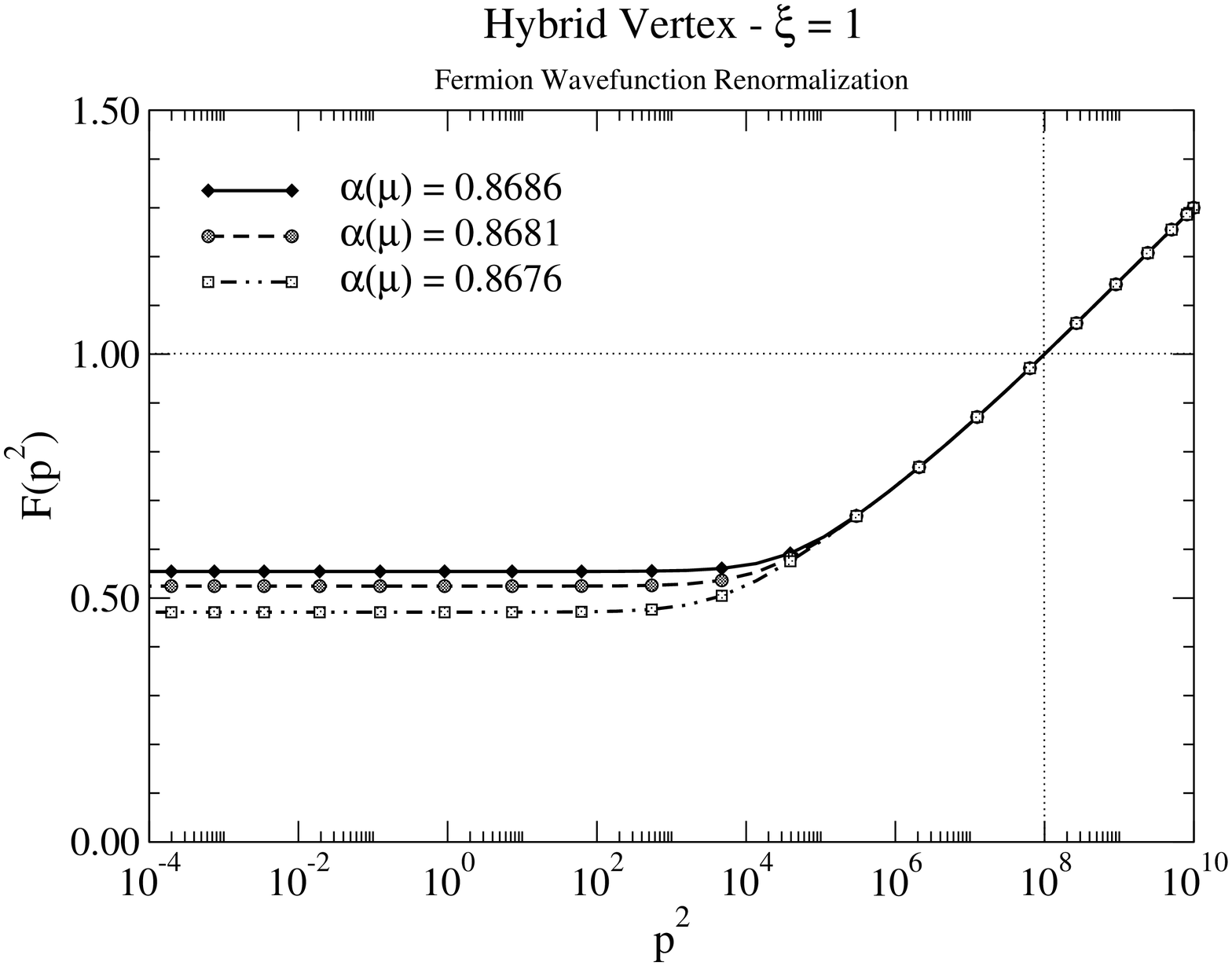}
\includegraphics[width=0.32\textwidth]{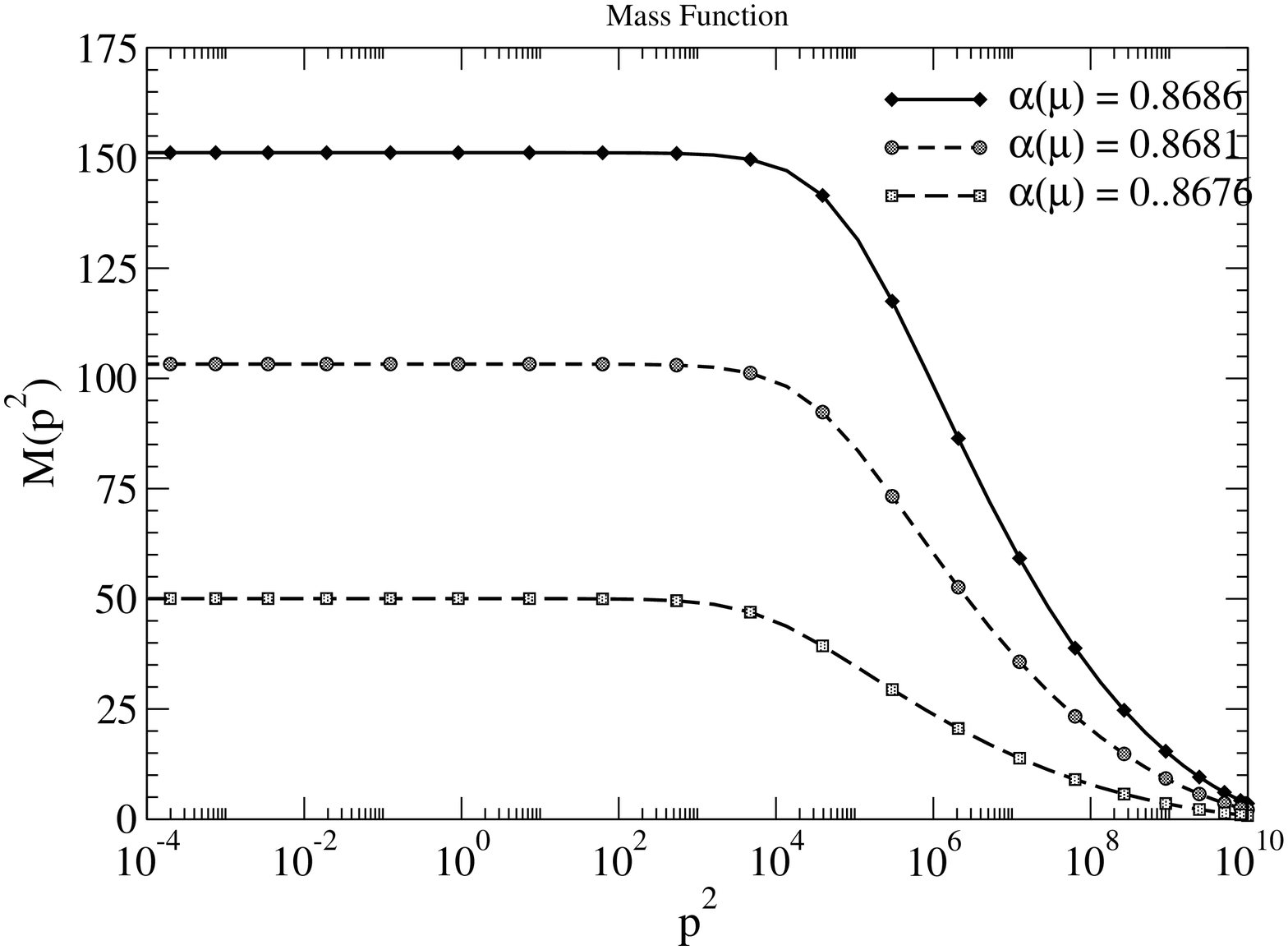}
\includegraphics[width=0.32\textwidth]{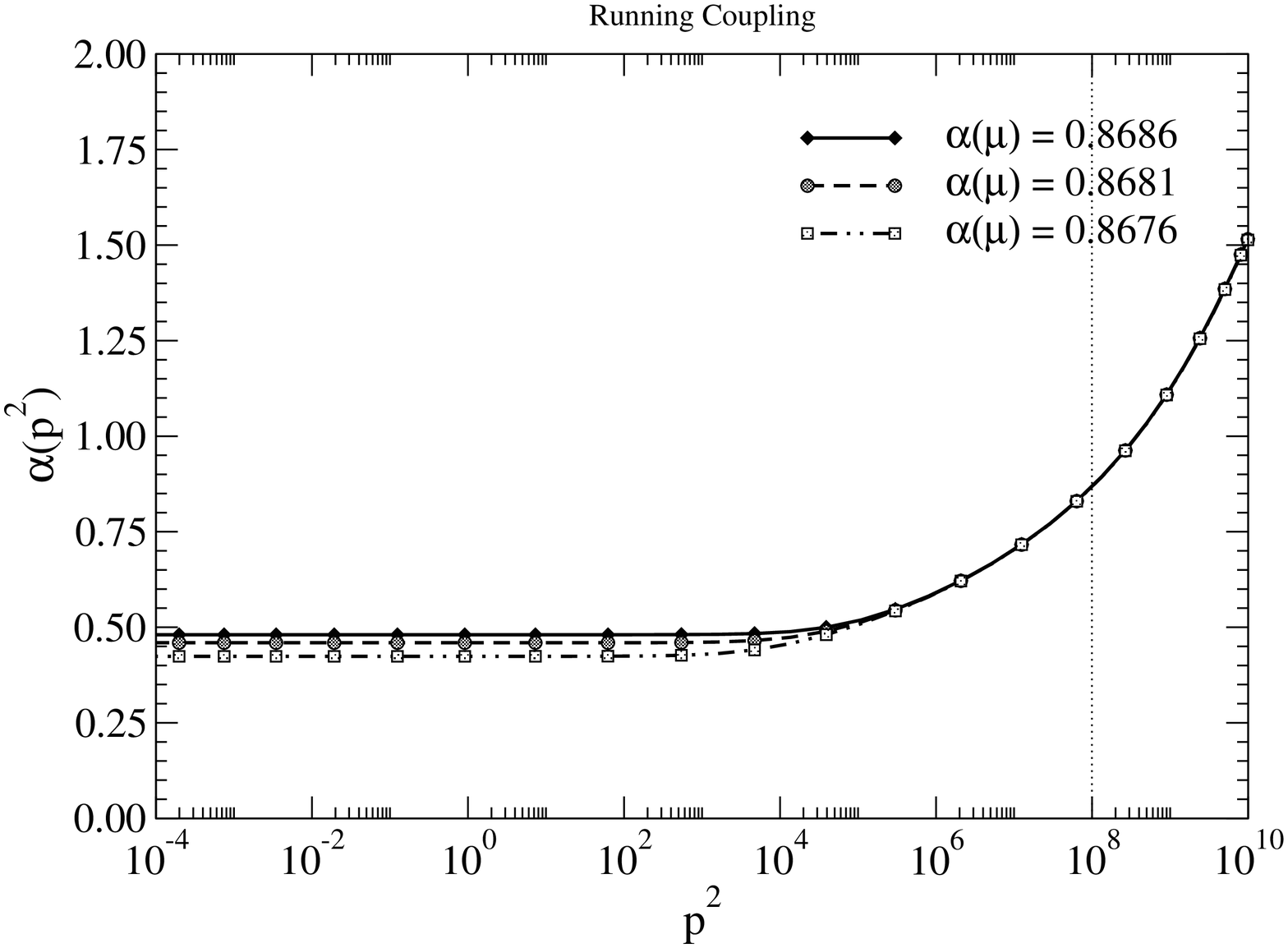}
}
\caption{Typical $m_{0} = 0$ solutions in the Landau gauge (top plots) and Feynman gauge (bottom plots), using the unquenched CP/KP hybrid vertex, for a symmetric momentum partition ($\eta = 1/2$). Shown, from left to right, are the wave-function renormalization, mass and running coupling (effective alpha) functions.}
\label{fig:Richard-hybridKP}
\end{figure*}

\begin{figure*}
\subfigure{
\includegraphics[width=0.32\textwidth]{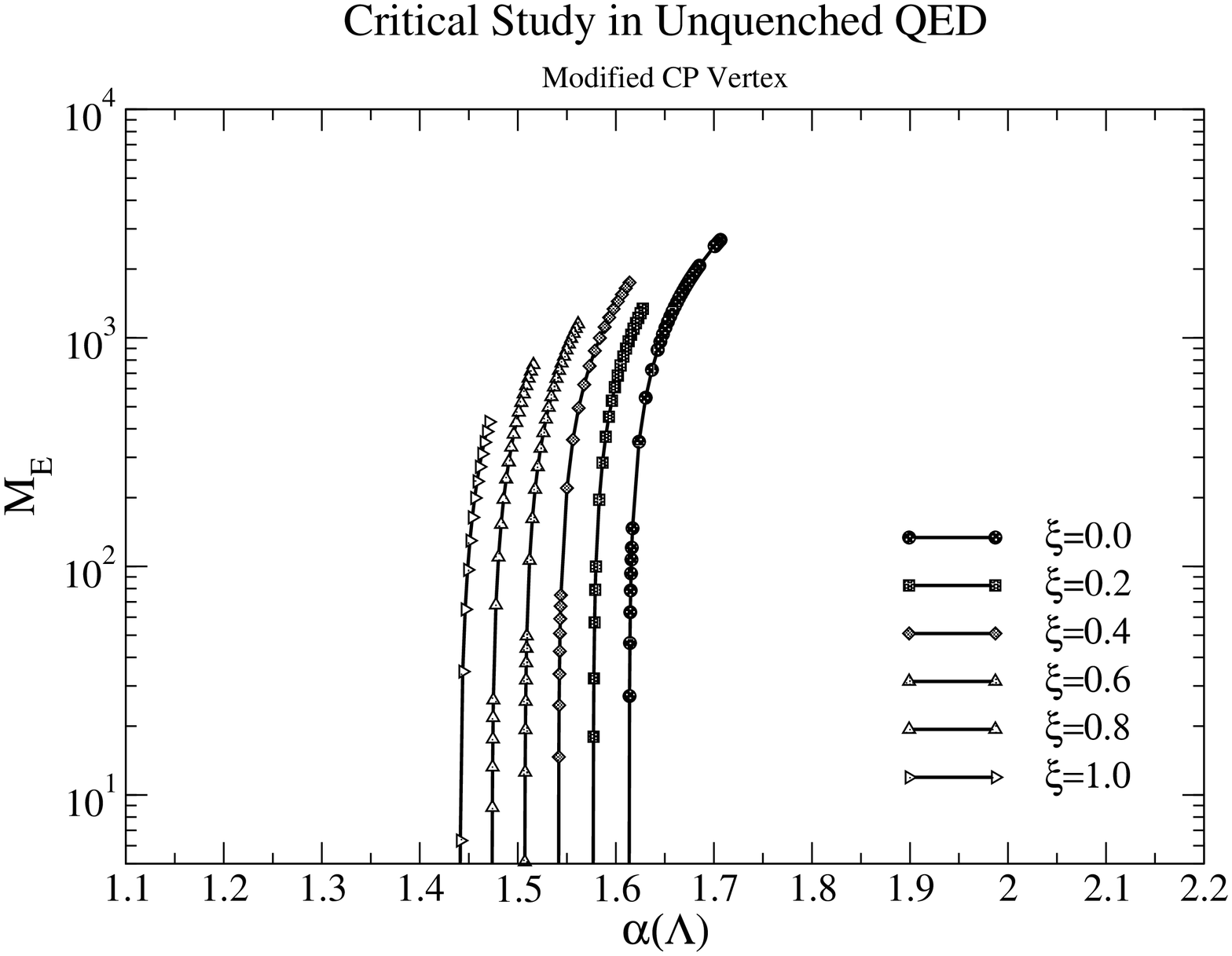}}
\subfigure{
\includegraphics[width=0.32\textwidth]{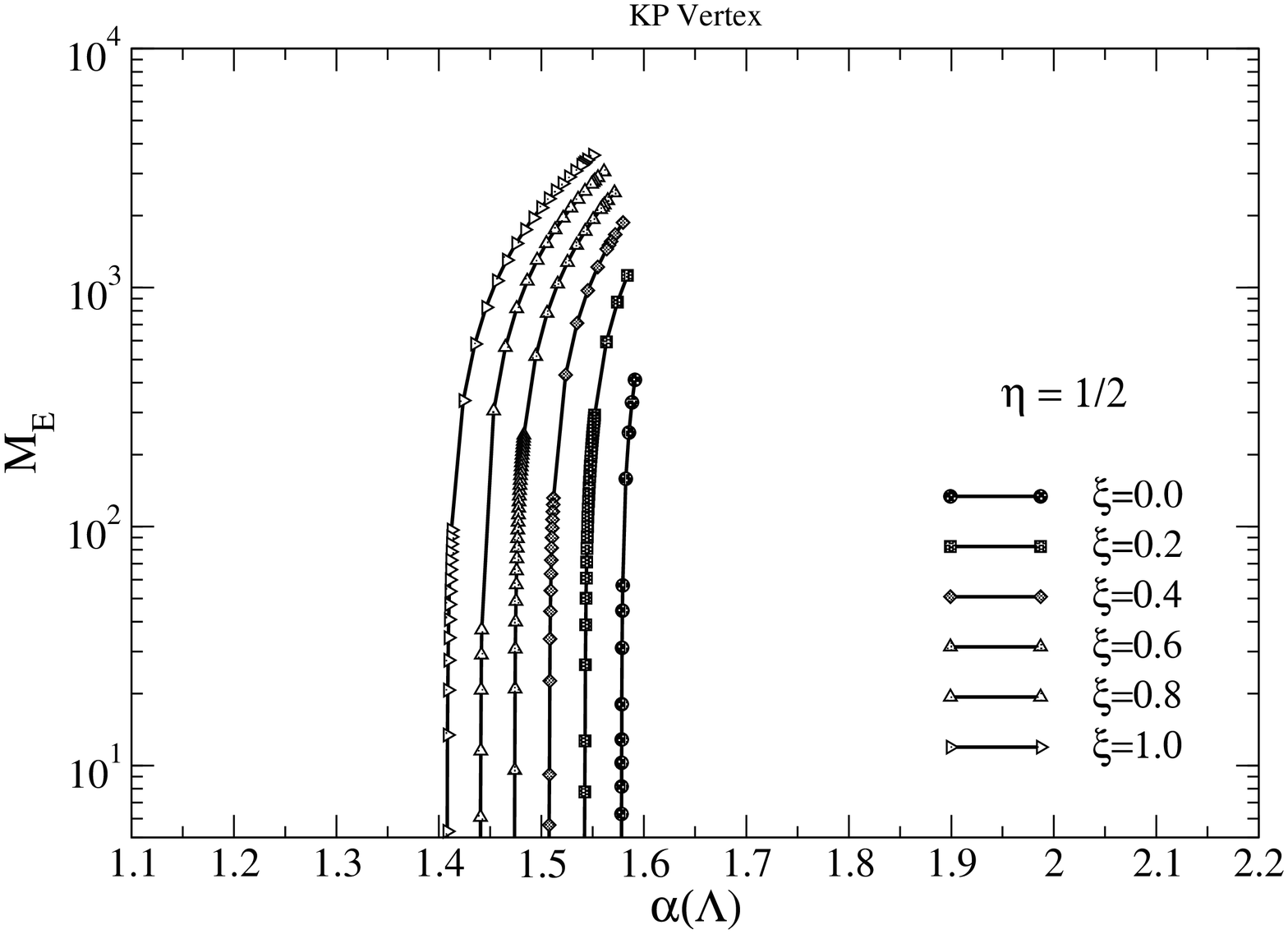}}
\subfigure{
\includegraphics[width=0.32\textwidth]{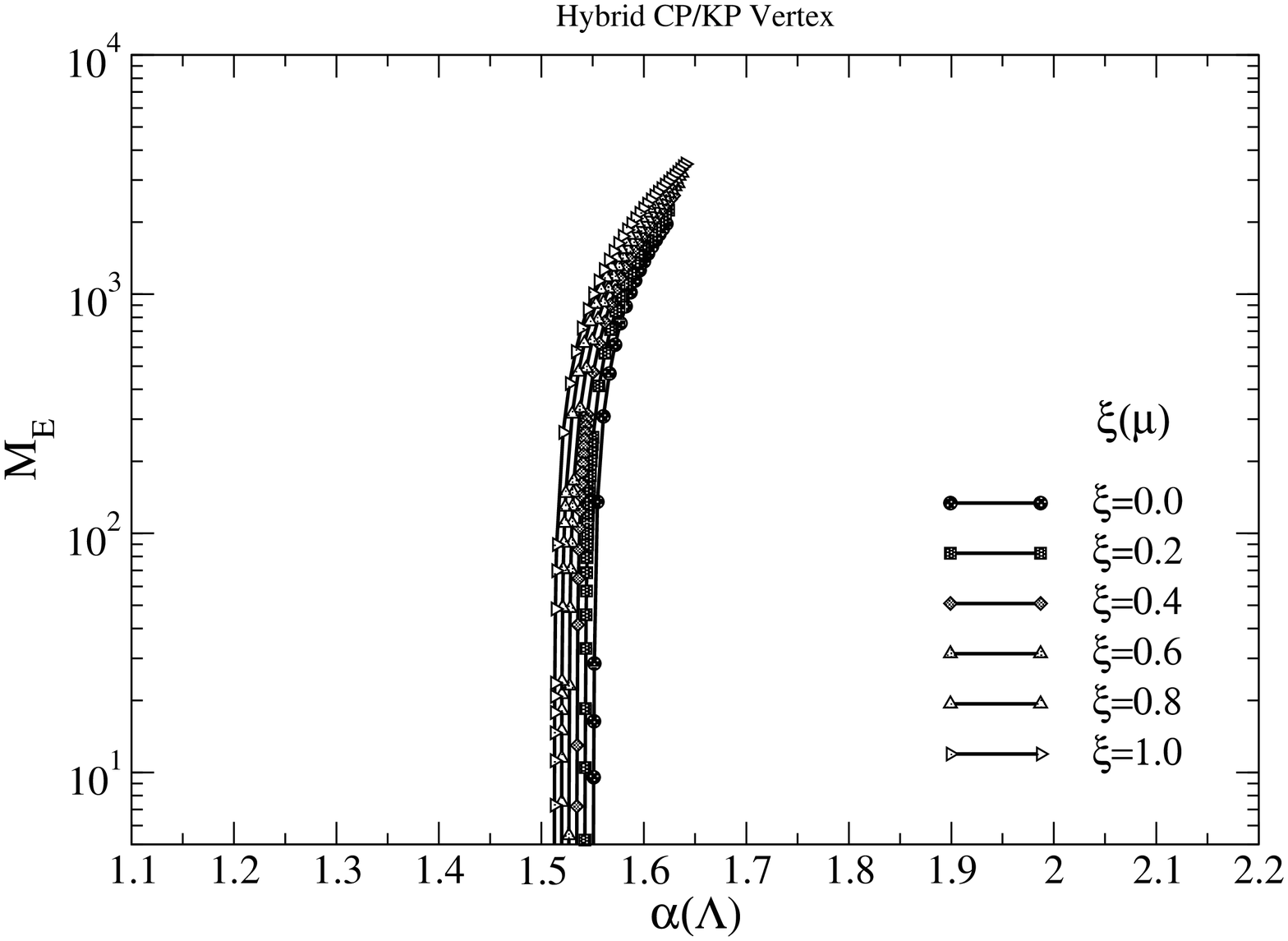}}
\caption{Euclidean mass vs.~coupling $\alpha_{\Lambda}$ for $m_{0}=0$ solutions for the modified~CP (left), KP (middle) and hybrid CP/KP (right) vertex Ans\"{a}tze for a spread of gauges and $\eta=1/2$. }
 \label{fig:richardmassgen}
\end{figure*}
\begin{figure*}[htp]
 \includegraphics[width=0.35\textwidth,angle=-90]{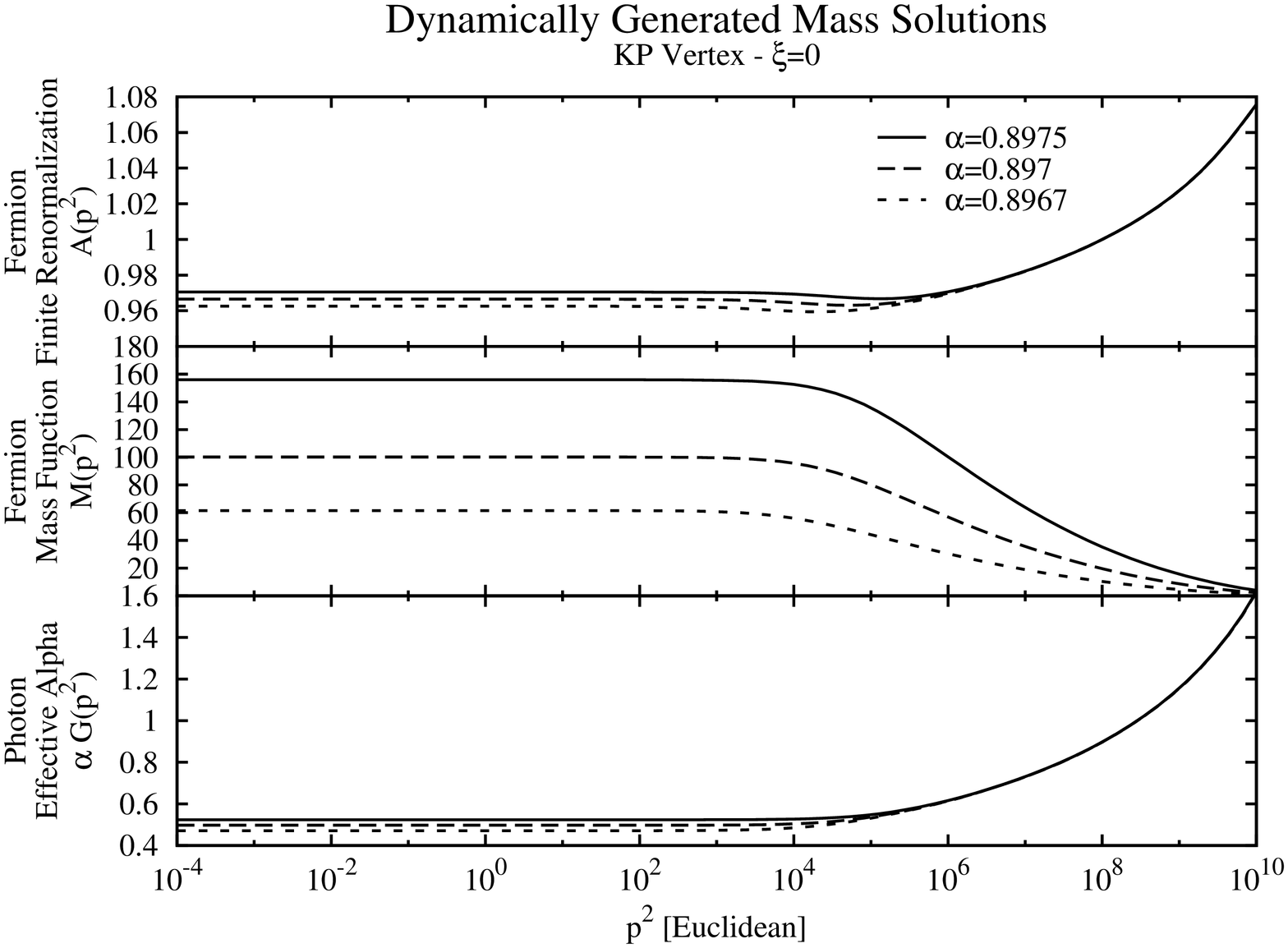}
 \includegraphics[width=0.35\textwidth,angle=-90]{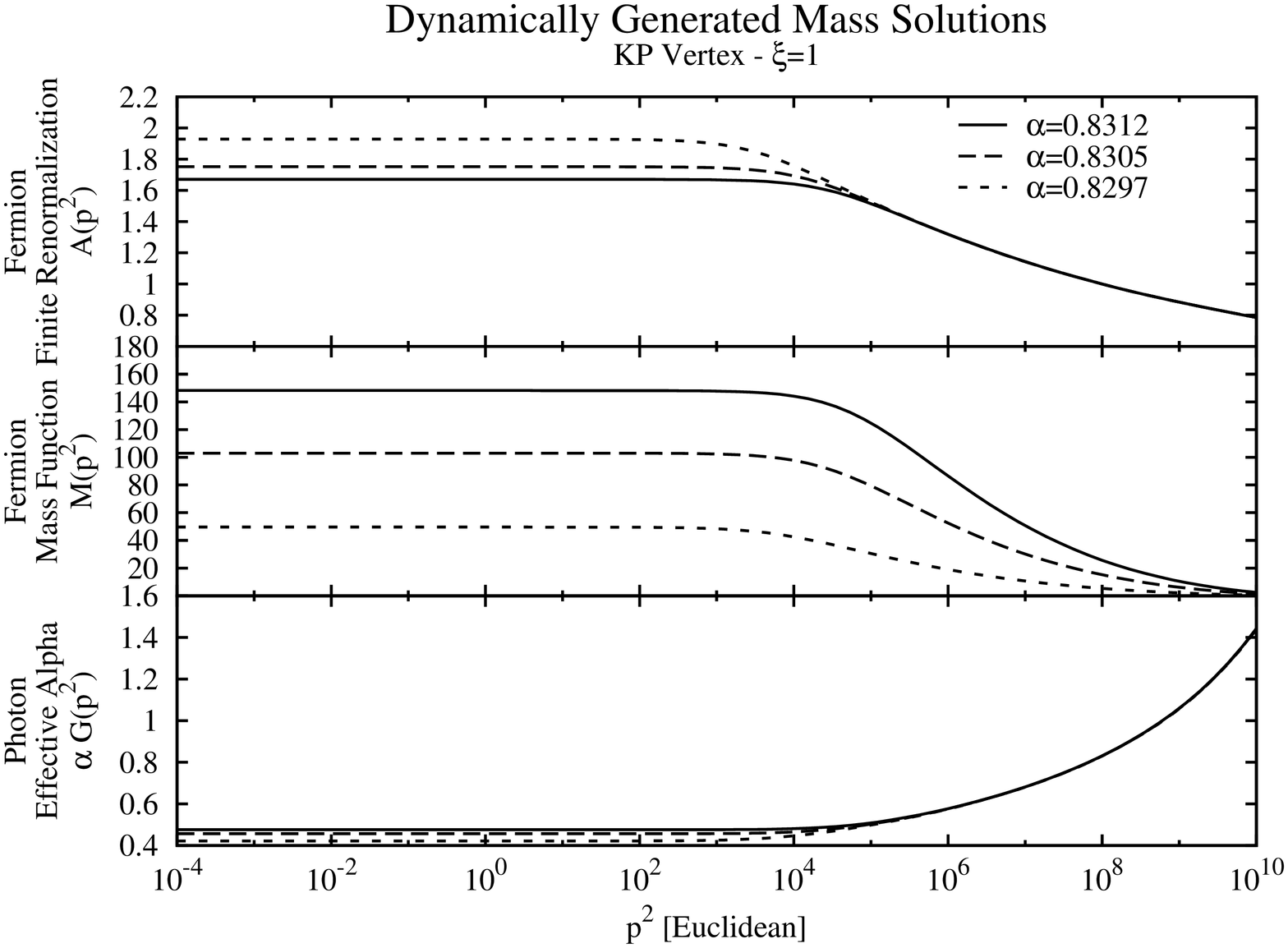}
 \caption{Inverse wave-function renormalization, mass-function and running coupling in Landau (left) and Feynman (right) gauges using the KP vertex in asymmetric momentum partition.}
 \label{fig:kp-chiral-alpha-feynman}
 \end{figure*}
\begin{figure}
\includegraphics[width=0.75\columnwidth,angle=-90]{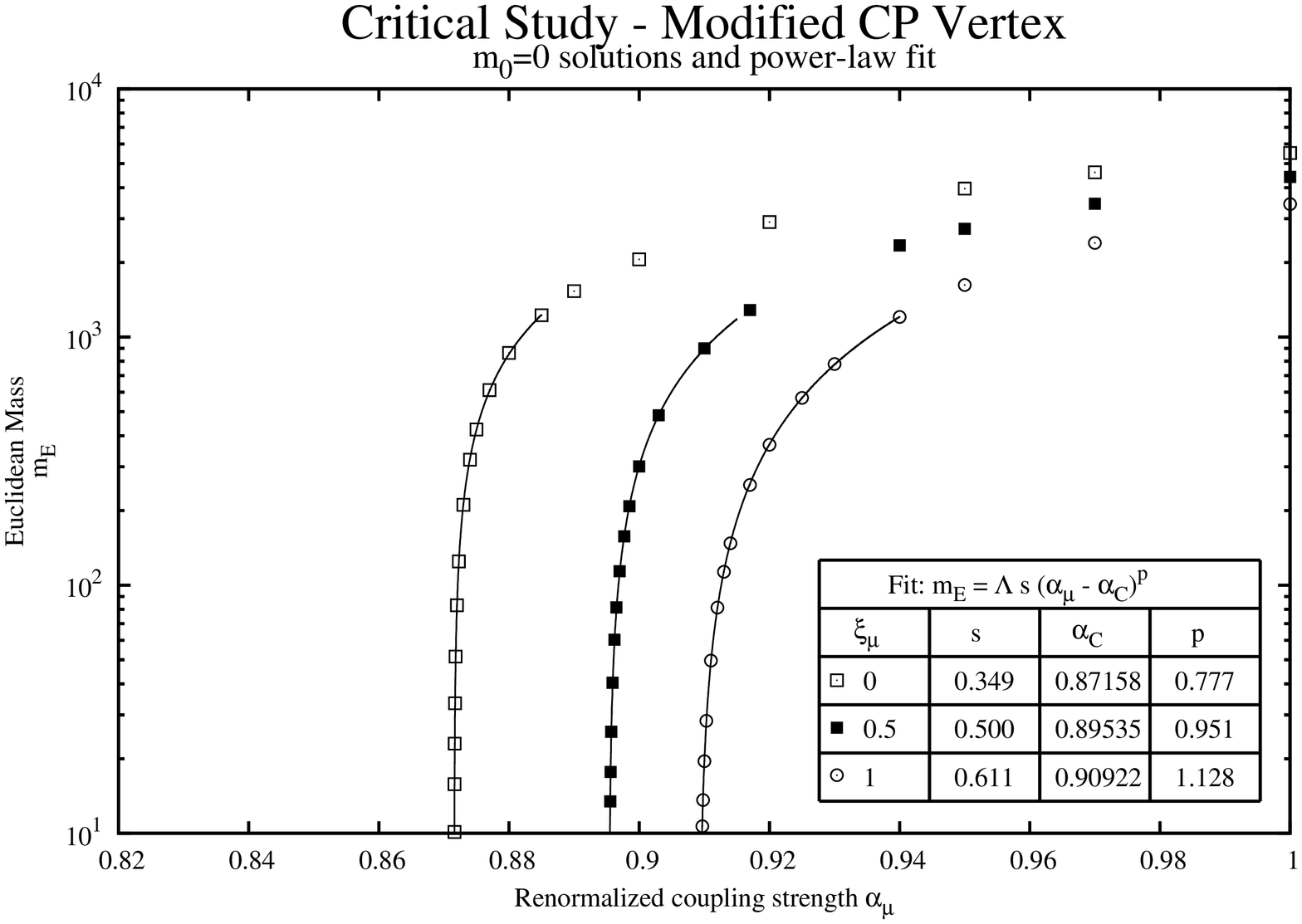}

\includegraphics[width=0.75\columnwidth,angle=-90]{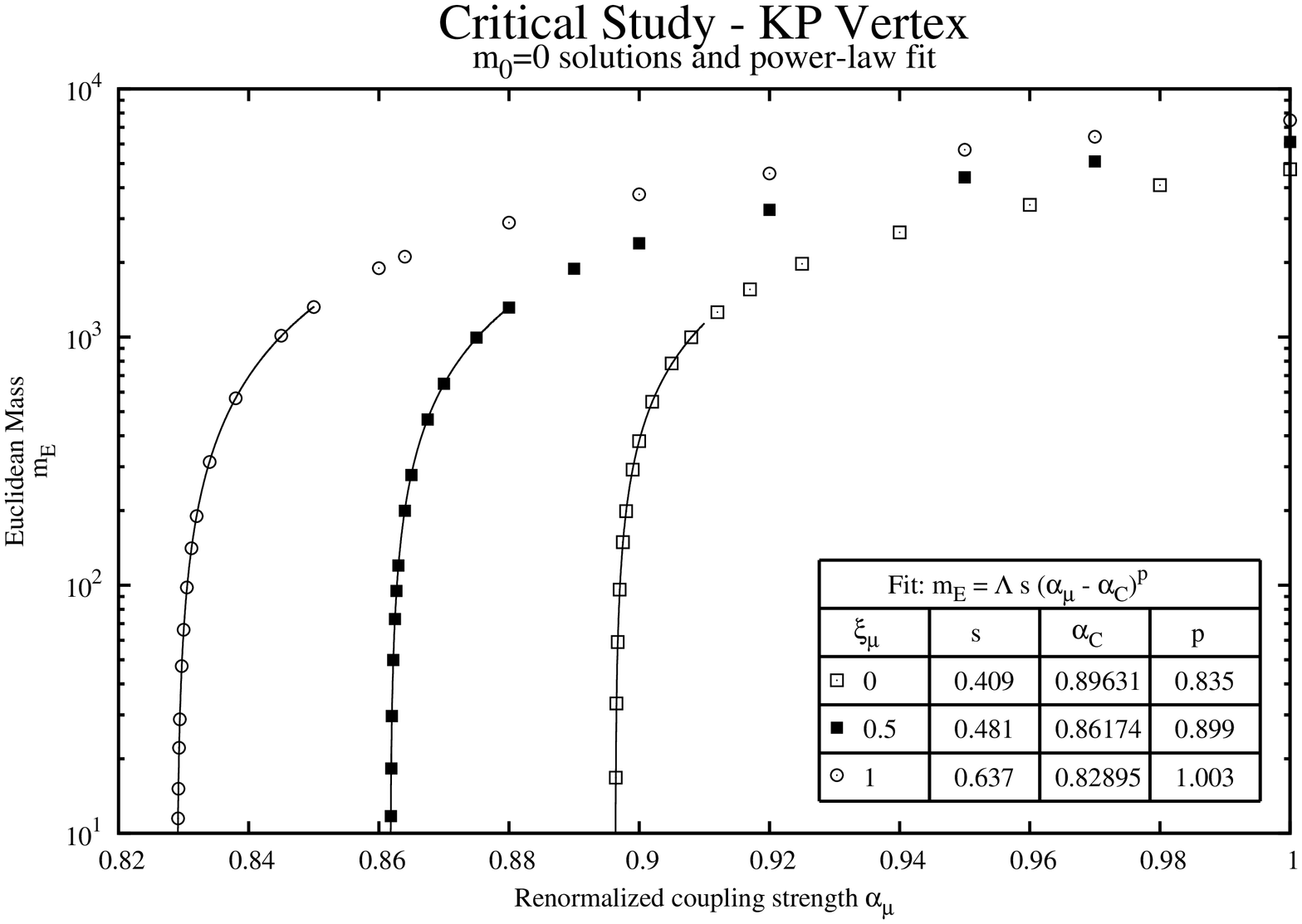}

\includegraphics[width=0.75\columnwidth,angle=-90]{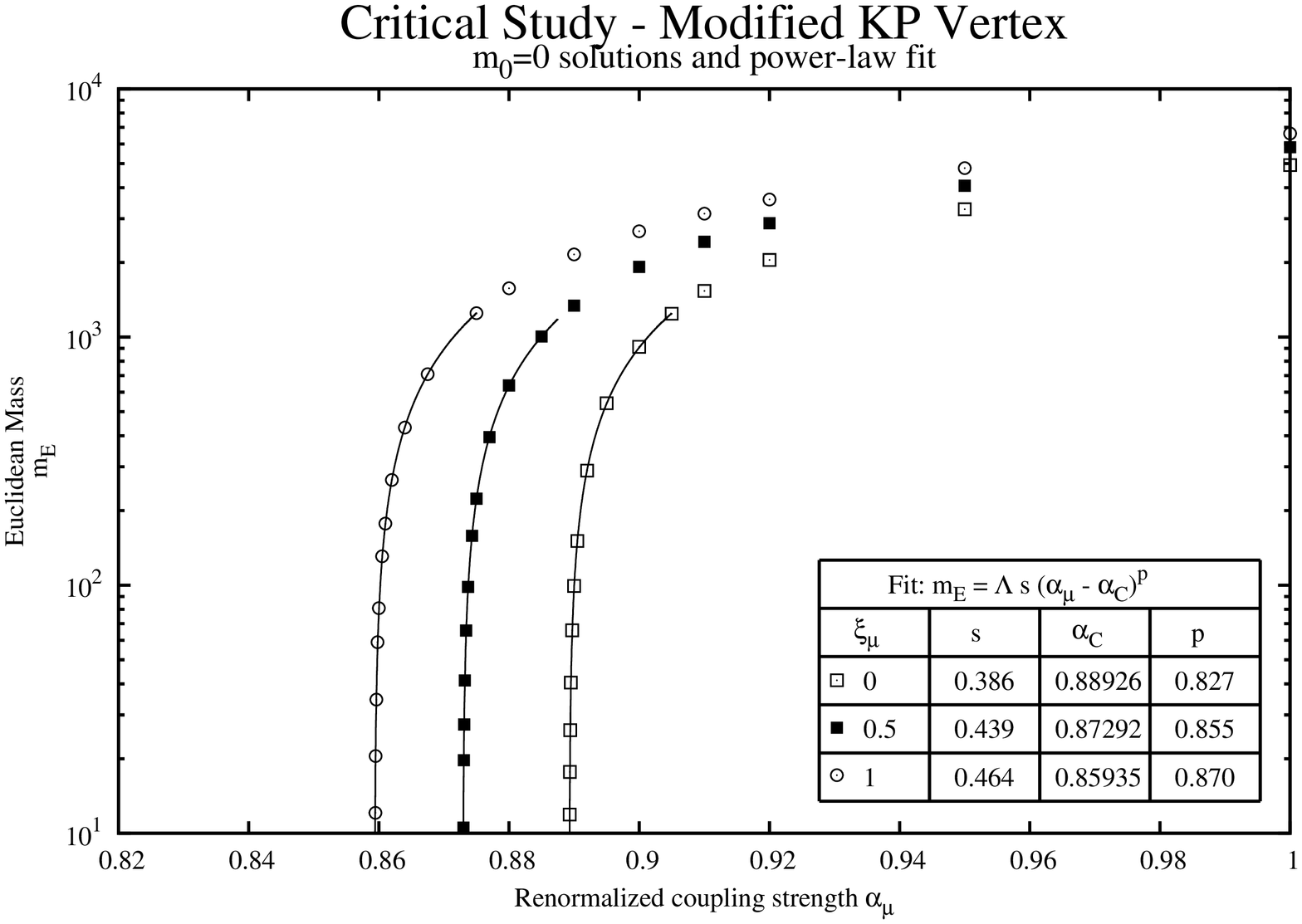}
\caption{Euclidean mass vs.~coupling $\alpha_{\mu}$ for $m_{0}=0$ solutions for the modified~CP (top), KP (middle) and hybrid CP/KP (bottom) vertex Ans\"{a}tze for $\xi_\mu=0,0.5,1$, $\eta=1$ and corresponding power-law fits in asymmetric momentum partition.}
\label{fig:mcp-massgen-cvg-fit}
\end{figure}

Figs.~\ref{fig:Richard-hybridKP} and \ref{fig:richardmassgen} illustrate the process for a symmetric momentum partition ($\eta=1/2$).
Each solution obtained at $\mu^{2} = 10^{8}$ for the hybrid CP/KP vertex in Fig.~\ref{fig:Richard-hybridKP} corresponds to an $(\alpha_\Lambda, m_{E})$ point in (the bottom) Fig.~\ref{fig:richardmassgen}, where $\alpha_{\Lambda} = \alpha_{eff}(\Lambda^{2})$ and $m_{E}$ is the Euclidean mass, derived from the mass function by the constraint
\begin{align}
	M(m_E^2) = m_E\,.
\end{align}
Here, and hereafter, solutions are renormalized at $\mu^{2}=10^{8}$ with $\Lambda^{2}=10^{10}$ and converged to one part in $10^5$ or better at each momentum point.

The critical coupling $\alpha_c$ is extracted from the solutions in Fig.~\ref{fig:richardmassgen} by a least-squares fit of the form
\begin{align}
	m_E = \,\Lambda\,s \, (\alpha_\Lambda - \alpha_c)^p
\end{align}
where $s$, $p$ and $\alpha_c$ are parameters to be fitted. Note that a simple power-law fit can be used, since the descent to criticality
is much steeper than in the quenched case. Of course, the critical coupling at the renormalization point, or the
unrenormalized critical coupling could be calculated instead, by fitting $(\alpha_\mu, m_{E})$ or $(\alpha_0, m_{E})$ points.
However, in all cases, the descent to criticality is along lines of constant $\xi_{\mu}$.

It is admitted that the consequent shift in $\xi$ explained in the previous section is a disadvantage 
to this method. However, in practice the shift is small (e.g. for KP $\xi=1$, the shift is $ \approx 1.7\%$) 
and decreases to insignificance as $\alpha \to \alpha_{c}$ reflecting the sharpness of the fall to criticality.

\begin{figure*}[htp]
\subfigure{\includegraphics[width=0.7\columnwidth,angle=-90]{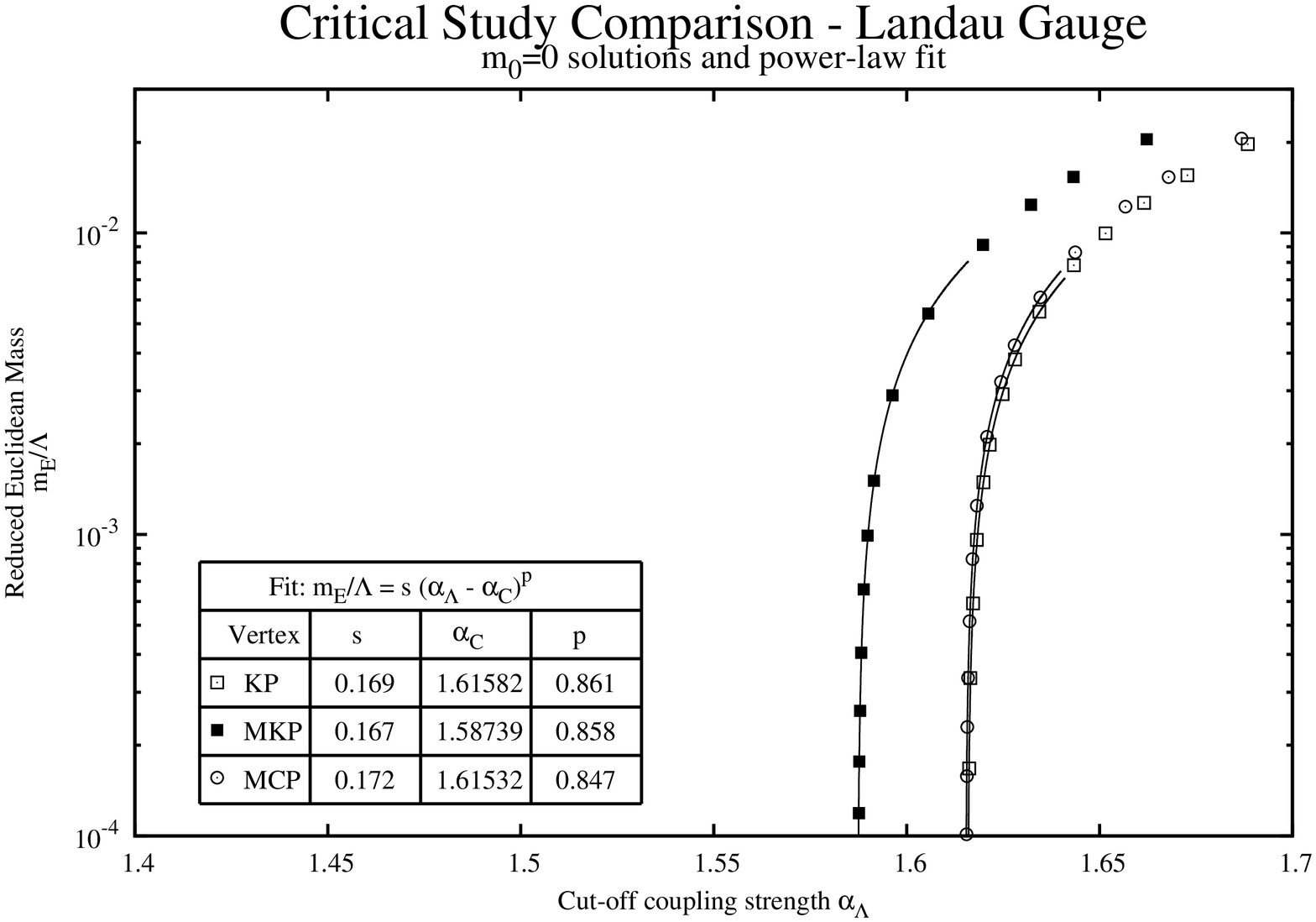}}
\subfigure{\includegraphics[width=0.7\columnwidth,angle=-90]{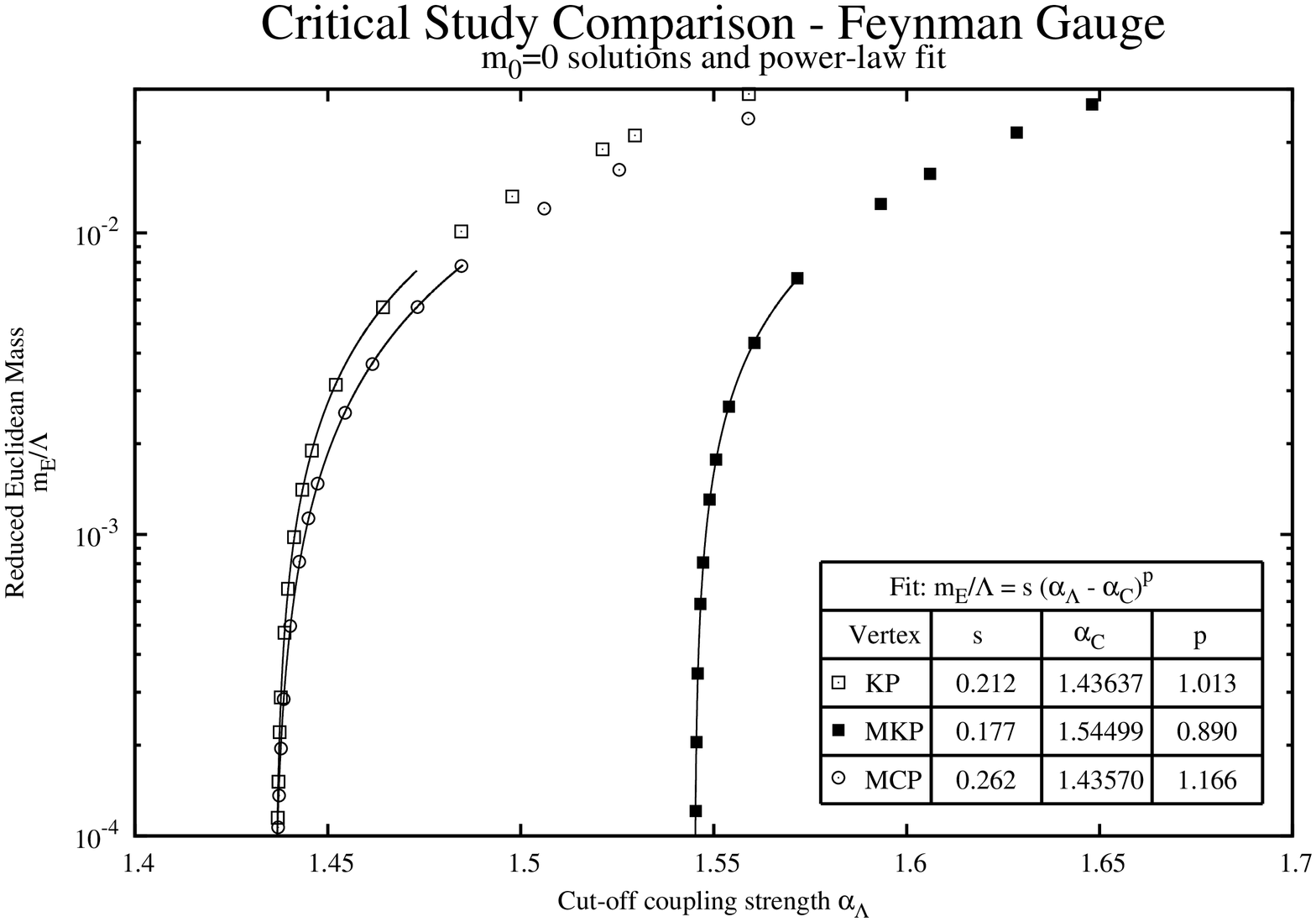}}\\
\caption{Vertex comparison of dynamically generated mass, $m_0=0$, solutions coupling at the cutoff, for the Landau (left)
 and Feynman (right) gauges and $\eta=1$.}
\label{fig:landau-massgen-cvg-fit-cutoff}
\end{figure*}

Figs.~\ref{fig:kp-chiral-alpha-feynman} and \ref{fig:mcp-massgen-cvg-fit} show results for the asymmetric momenta split ($\eta=1$), but for $\alpha_{\mu}$;
the results for $\alpha_{\Lambda}$ are shown in Fig.~\ref{fig:landau-massgen-cvg-fit-cutoff} which compares vertex Ans\"{a}tze
in the Landau (top figure) and Feynman (bottom figure) gauges. The outcomes for the Landau and Feynman gauges in both momentum partitioning
schemes are summarized in Table~\ref{table:mcp-massgen-alphas}: we adopt the expedient of truncating the shifted $\alpha_{\Lambda}$ and
$\xi_{\Lambda}$ results to four and two decimal places respectively.

We note that there is a small but significant variation in the results for the different momentum partitioning.
We further note that in all cases, and using either momentum partitioning scheme,
the hybrid CP/KP vertex exhibits the least gauge variance. Also, for $\alpha_{c}$ calculated at the cutoff, the modified CP and KP vertices
give very similar results in either momentum partitioning scheme for both Landau and Feynman gauges, although 
there is a wide gauge variation. However, remarkably, the results for $\alpha_{c}$ at the renormalization point differ widely for the modified CP and KP vertices.

\begin{table*}
\caption{Critical couplings at the renormalization point ($\alpha_{\mu}$), at the cut-off ($\alpha_{\Lambda}$)
 and unrenormalized ($\alpha_{0}$) for
the Landau and Feynman gauges using the modified CP vertex, KP vertex and hybrid CP/KP vertex,
for symmetric and  asymmetric momentum partitions.
Note that, except in Landau gauge, calculating $\alpha_{\Lambda}$ and $\alpha_{0}$ necessitates a shift in $\xi$ as well.
These results were generated by the Durham group~\cite{Williams:2007zzh}. The asymmetric results were also
generated by the Adelaide group independently and found to be in agreement.}
\label{table:mcp-massgen-alphas}
\begin{ruledtabular}
\begin{tabular}{c|c||c|c|c||c|c|c|c|c}
\multicolumn{2}{c||}{} &
\multicolumn{3}{c||}{Symmetric ($\eta=1/2$)} &
\multicolumn{5}{c}{Asymmetric ($\eta=1$)} \\
\hline
\hline
Vertex     &  \,\, $\xi_\mu$ \,\,& \,\, $\alpha_\mu$ \,\,& \,\,$\xi_\Lambda$ \,\,& \,\,$\alpha_\Lambda$ & \,\, $\alpha_\mu$ \,\,& \,\,$\xi_\Lambda$ \,\,& \,\,$\alpha_\Lambda$ \,\,& \,\,$\xi_0$\,\, &\,\, $\alpha_0$ \\
\hline
Mod. CP    &$0    $ & $0.87127    $ & $0.00             $ & $1.6135 $ & $0.87158    $ & $0.00             $ & $1.6152                 $ & $0.00$ & $1.6001     $ \\
Mod. CP    &$1    $ & $0.90681    $ & $0.63             $ & $1.4409 $ & $0.90921    $ & $0.63             $ & $1.4358                 $ & $0.62$ & $1.4749     $ \\
\hline
\hline
KP         & $0   $ & $0.90567    $ & $0.00             $ & $1.5783 $ & $0.89632    $ & $0.00             $ & $1.6158                 $ & $0.00$ & $1.5989     $ \\
KP         & $1   $ & $0.83658    $ & $0.59             $ & $1.4080 $ & $0.82895    $ & $0.58             $ & $1.4361                 $ & $0.56$ & $1.4766     $ \\
\hline
\hline
Hybrid KP  & $0   $ & $0.89860	  $ & $0.00             $ & $1.5504 $ & $0.88926    $ & $0.00             $ & $1.5873                 $ & $0.00$ & $1.5701     $ \\
Hybrid KP  &$1    $ & $0.86726	  $ & $0.57             $ & $1.5126 $ & $0.85935    $ & $0.56             $ & $1.5449                 $ & $0.54$ & $1.5944     $ \\
\end{tabular}
\end{ruledtabular}
\end{table*}

It is evident that despite the successes of the KP vertex in the massless limit, where the breaking of gauge-invariance was
significantly suppressed by the satisfying of multiplicative renormalizability, it is not as good in this regard as the
hybrid CP/BC vertex when dynamical mass generation is manifest. Presumably, the reason for this is that the KP vertex does not
yet include any mass terms in it, i.e. the transverse form factors $\tau_{1,4,5,7}$ in the transverse vertex,
Eq.~(\ref{eq:transverse}), have been chosen to vanish.
We envisage that this means that the leading and sub-leading logarithms in a perturbative expansion of the mass function
are not correctly related.

\subsection{\label{sec:Condensate}Condensate}
\begin{figure}
\includegraphics[width=0.7\columnwidth,angle=-90]{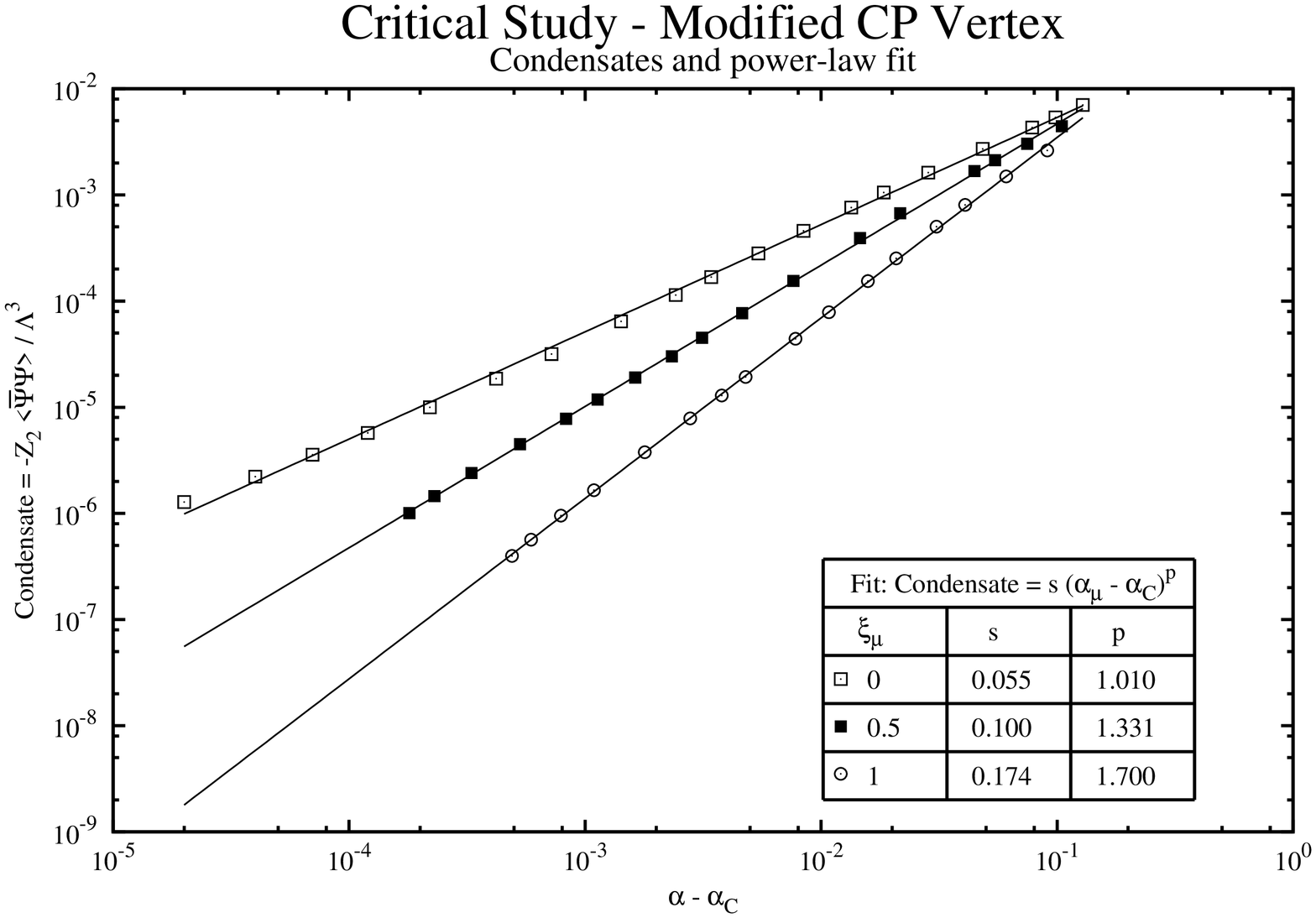} 

\includegraphics[width=0.7\columnwidth,angle=-90]{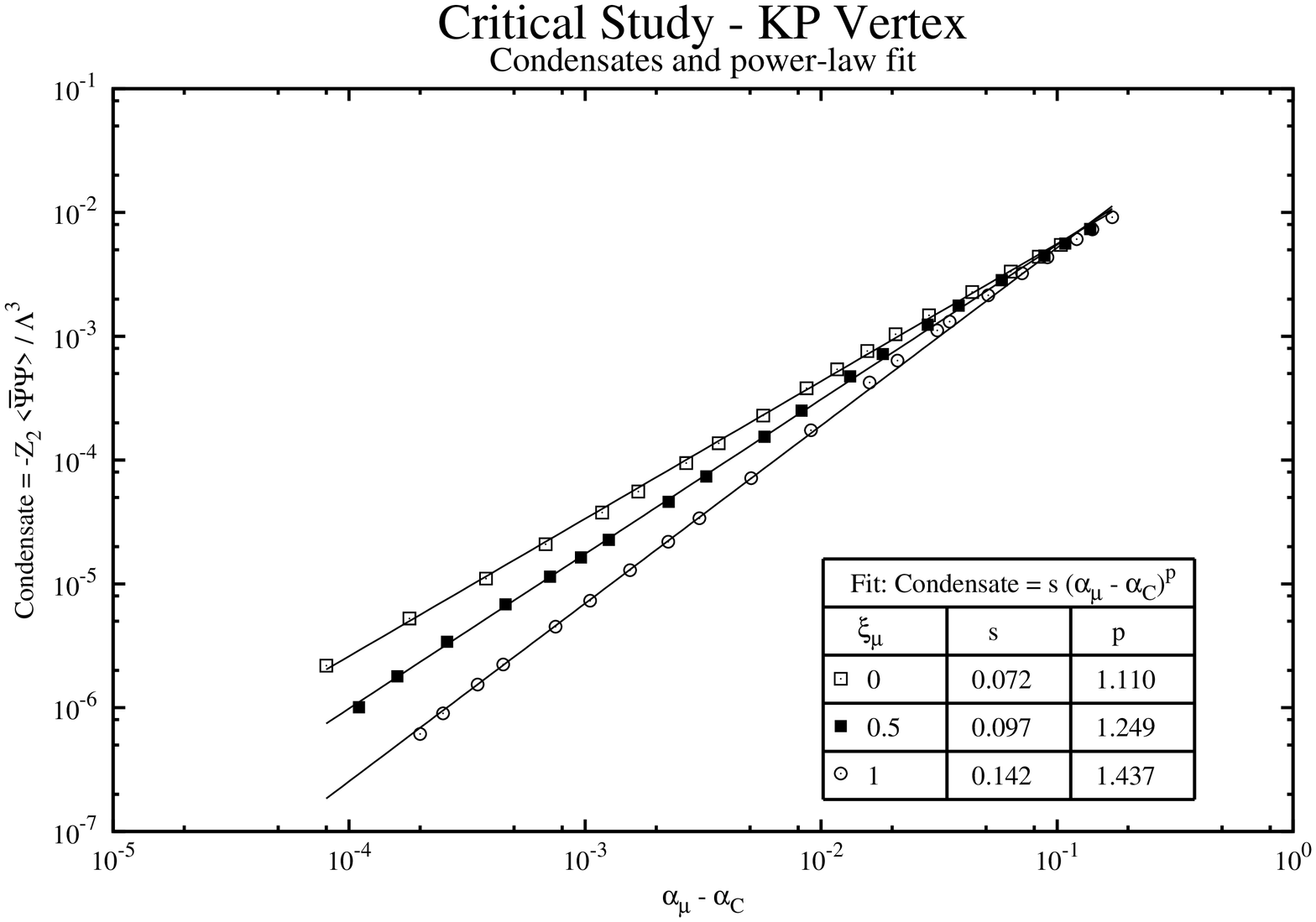} 

\includegraphics[width=0.7\columnwidth,angle=-90]{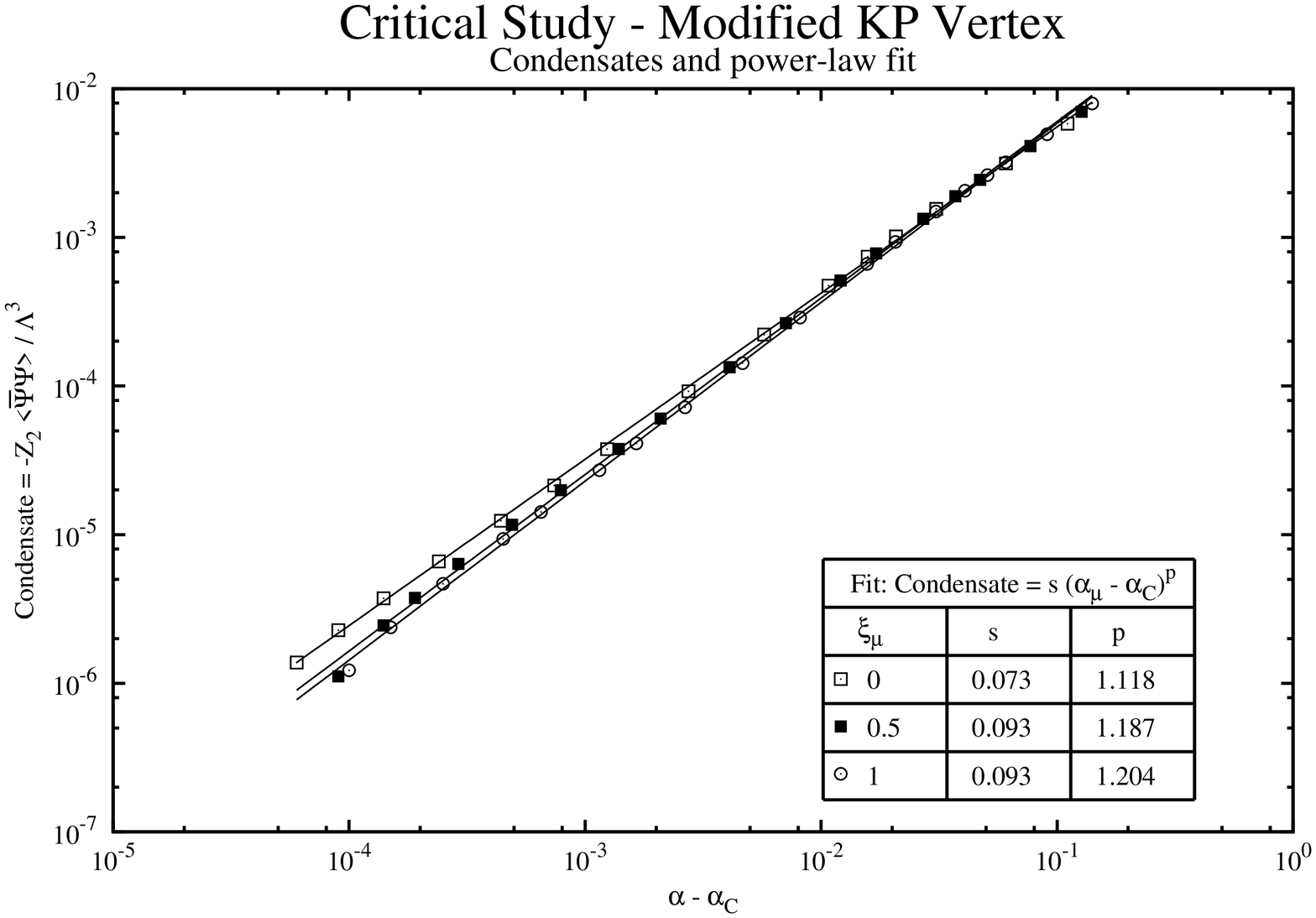}
\caption{A log-log plot of the condensate versus $\alpha-\alpha_c$ for $m_0=0$ using modified~CP vertex (top), KP vertex (middle) and hybrid CP/KP vertex (bottom).}
\label{fig:condensate}
\end{figure}

Another signal of dynamical mass generation is that the condensate
\begin{align}
	\langle 0|{\bar{\Psi}}\Psi | 0 \rangle = -\frac{4}{\pi} \, \int dp^2 \frac{p^2\,B(p^2)}{p^2\,A^2(p^2) + B^2(p^2)}\;,
\end{align}
is non-zero if and only if we are above criticality.
In Fig.~\ref{fig:condensate}, we show the condensates for the solutions in Fig.~\ref{fig:mcp-massgen-cvg-fit}, as functions of
$\alpha-\alpha_c$, and resultant fits which show power-law behaviour. Also notable is the condensates exhibit the same gauge-variant behaviour 
as the solutions in Fig.~\ref{fig:mcp-massgen-cvg-fit}: that is, the hybrid CP/KP vertex solutions exhibit the least gauge variance,
and the KP vertex the most.

\subsection{Comparison With Other Studies}
Ref.~\cite{Bloch:1995dd} uses the modified~CP vertex in their unquenched QED$_4$ work where they solve the DSEs by iteration. They found 
the critical coupling $\alpha_c(\Lambda^2, N_F=1) = 1.61988 $ which agrees with our results $\alpha_c(\Lambda^2, N_F=1)=1.61567$
in Landau gauge very well.
Bashir et.al~\cite{Bashir:2011dp} studied dynamical mass generation in Landau gauge using the KP vertex together with another additional transverse piece for various number of fermion flavours their result for the critical coupling $\alpha_c^A(N_F=1) = 2.27$ (analytically and $\Lambda^2 \rightarrow \infty$ is taken) and  $\alpha_c^N(\Lambda^2,N_F=1) = 2.4590$ (numerically) and they found their analytical and numerical results on the dynamically generated mass function differs slightly. Both the solutions exhibits the scaling law, 
$M(0)/\Lambda^2 = a\,(\alpha-\alpha_c)^p$ with the exponent of $p=0.5$ (analytically), $p=0.7819$ (numerically).
Our critical value for $N_F=1$ is much smaller than theirs but our exponent of the power law $p= 0.89631$ is larger than theirs.
Referring again to Fig.~\ref{fig:Richard-hybridKP}
for comparison, we chose values such that the resulting masses $M(0)$
were approximately equivalent. Clearly, the wave-functions show their intrinsic gauge variance. This is present also
in the mass-functions, but is masked somewhat by the generation of a mass and is mainly discernible from differences in
the ultraviolet behaviour. The key test, however, is in the dressing functions for the photon propagator which
should be invariant under a change of gauge. We see that the solutions for two gauges are in fact very similar -- as is
most evident from the plots of Fig. \ref{fig:mcp-massgen-cvg-fit}.

\section{\label{sec:Conclusion} Conclusion and Future Work}

In this article we studied dynamical mass generation in unquenched QED$_{4}$ by solving the DSEs for the fermion
and photon propagators. Given a suitable Ansatz for the fermion-photon vertex, the infinite tower of equations can be truncated, forming
a closed system of coupled non-linear integral equations that are solved numerically by iteration.

In the absence of any explicit mass term in the Lagrangian, fermions remain massless until the gauge coupling reaches some critical value
whereupon the fermions acquire a dynamical mass. This occurs in both quenched QED (where the coupling does not run and the photon propagator
is trivial) and unquenched QED. The value of the critical point depends on the choice of vertex adopted to solve these equations. Ideally,
the photon propagator and critical coupling should be independent of the choice of gauge: its actual degree of gauge variance is an
important constraint for constructing a fermion-photon vertex as well as a test for checking the quality of the truncation introduced.
While the the bare vertex leads to a highly gauge dependent critical coupling, the CP and KP vertices perform much better in this regard due to their construction.
Recall the CP vertex was constructed to make the fermion propagator consistent with multiplicative renormalizability. Since this represents a kinematic region in which a large momentum flows through a fermion line to the photon leaving the other fermion with fixed momentum, it fails in the photon equation when ultra-violet renormalization concerns large momenta in the fermion legs, with the photon momentum fixed. Consequently, the CP vertex, widely used in quenched studies, is incompatible with unquenched QED, except in the massless case. In contrast, the KP vertex aims to make both the fermion and photon propagator equations consistent with multiplicative renormalizability in the case of massless fermions.

The main aim of this paper has been to explore the performance of the KP vertex with regard to gauge invariance when a
dynamical mass is generated. This analysis has highlighted the sensitivity to the fact that the design of the KP vertex is  incomplete, since no mass terms are included in its structure. Hence, for comparison purposes,
two hybrid vertices were also considered: these used the CP vertex for the fermion equation, and the Ball-Chiu and the KP vertex
respectively for the photon equation. It was found that the latter vertex (the hybrid CP/KP vertex) exhibited the least gauge variance.
Another possible source of gauge variance is the use of a cutoff regulator: by employing two different schemes for evaluating the fermion
loop momenta, it was found that this effect was also non-trivial.  This illustrates how our understanding of strong coupling QED is as yet incomplete, and in need of further study.

As soon as   fermion loops are introduced in the QED vacuum, the electric charge is  screened and the effective coupling starts to run. One of the outstanding questions in full QED is how much screening the effective running coupling will receive in the continuum limit $(\Lambda \longrightarrow \infty)$? Will the effective running coupling die off and the theory  become free and non-interacting? Or at large momenta does the coupling become so big that new operators, like that of four-fermions, become relevant in such a  way that the theory remains interacting, and non-trivial. 
Answering these questions is outside the scope of this paper, but the subject of subsequent work to be reported elsewhere. There the relevant operators will be added to the Lagrangian allowing their effect to be quantified.

\begin{acknowledgments}
%
%
We would also like to thank C.\ S.\ Fischer and C.\ D.\ Roberts for useful discussions. AK and TS thank \ A.\ W. Thomas 
for supporting this study under the aegis of the Centre for the Subatomic Study of Matter (CSSM). We also acknowledge support 
from the Australian Research Council International Linkage Award (LX 0776452), the Australian Research Council Discovery 
grant (DP0558878), and the Austrian Science Fund (FWF) under project number M1333-N16. MRP acknowledges support of Jefferson Science Associates, LLC under U.S. DOE Contract No. DE-AC05-06OR23177 for the writing of this paper.
\end{acknowledgments}

\appendix

\section{\label{sec:DSE}Dyson--Schwinger equations (DSEs)}

The fermion and the photon Dyson--Schwinger equations which are solved iteratively for the 
fermion and photon wave function renormalizations and for the mass function are given below. 

\subsection{\label{subsec:FermionDSE}Fermion Wave-function Renormalization}

The fermion self-energy in Eq.~(\ref{eq:mainsdf}) can be decomposed into Dirac and scalar terms,
\noindent
$\protect{\overline{\Sigma}(p)= \overline{\Sigma}_d(p)\,{\not\!p}+{\overline{\Sigma}}_s(p)}$ which is obtained from $\Sigma(p)$ by

\begin{align}
{\overline{\Sigma}}_d(p^2)=\frac{1}{4}\,{\rm Tr} \left(\overline{\Sigma}(p)\,\frac{\not\! p}{p^2}\right),  \qquad
{\overline{\Sigma}}_s(p^2)=\frac{1}{4}\,{\rm Tr} \left(\overline{\Sigma}(p)\cdot \mathbbm{1}\right)\,.
\label{eq:sigmas}
\end{align}

\noindent
Multiplying Eq.~(\ref{eq:mainsdf}) by $\not\!p$ and $\mathbbm{1}$ respectively yields two separate
equations for the inverse fermion wave-function renormalization and the mass function~:
\begin{align}
F^{-1}(\mu^2;p^2)         &=  Z_2(\mu)\phantom{\; m_0}      \;- \; Z_2(\mu) \;\overline{\Sigma}_d(p^2)\,,
\label{eq:zeq}\\
M(p^2)\; F^{-1}(\mu^2;p^2)&=  Z_2(\mu)\; m_0\; +\; Z_2(\mu)\>\overline{ \Sigma}_s(p^2)\,.
\label{eq:meq}
\end{align}
Evaluating Eqs.~(\ref{eq:zeq},\ref{eq:meq}) at the renormalization point, $p^2=\mu^2$,
and forming an appropriate difference one can eliminate the divergent constants 
$Z_1$ and $Z_2$ to obtain the renormalized quantities
\begin{widetext}
\begin{align}
F(\mu^2;p^2) &=  1 \> + \> F(\mu^2;p^2) \>\overline{ \Sigma}_d(p^2) \> - \> \overline{ \Sigma}_d(\mu^2)\,, 
\label{eq:ferwavefunc} \\
\overline{\Sigma}_d(p^2)&=  \frac{\alpha}{4\pi^3}\, \int_E d^4k\,\frac{1}{p^2} \frac{1}{q^2}\,\frac{F_k}{
\left[k^2+M_k^2\right])}\,\Bigg[ 
 - \frac{\xi}{q^2} \frac{1}{F_p}\, \left[  p^2\, k \cdot q +M_k M_p\,p \cdot q \right] \nonumber\\
 & +  \frac{G_q}{q^2} \Bigg\{\,\,\,\,\,\,\,\,\,
                            \frac{1}{2}\,    \left(\frac{1}{F_k}+\frac{1}{F_p}\,\right)\, \left[-2\Delta^2-3q^2 k \cdot p  \right]
                            +\frac{1}{2\,(k^2-p^2)} \left(\frac{1}{F_k}-\frac{1}{F_p}\right) \left[ -2 \Delta^2 (k^2+p^2) \right] \nonumber\\
                  & \hspace{20mm} + \frac{1}{(k^2-p^2)}    \left(\frac{M_k^2}{F_k}-\frac{M_p\,M_k}{F_p}\right)\, \left[ - 2\Delta^2\right]
                              \Bigg\}\, \nonumber\\ 
&  +
G_q \Bigg\{\,\,\,\,\,\,\,\,\,
                                       \tau_2^E(p^2,k^2,q^2) \left[-\Delta^2 (k^2+p^2) \right]  + \tau_3^E(p^2,k^2,q^2) \left[2\Delta^2 +3q^2 k \cdot p \right] \nonumber\\
  &   \hspace{20mm} + \tau_6^E(p^2,k^2,q^2) \left[3 \, k \cdot p \,(p^2-k^2) \right] + \tau_8^E(p^2,k^2,q^2) M_k\left[-2 \Delta^2 \right]\quad \Bigg\}\,\,\,
\Bigg] \,,
\label{eqn:dirac_self_energy}
\end{align}
\end{widetext}
where $\Delta^2=(k \cdot p)^2-k^2p^2$ and 
\begin{widetext}
\begin{align} 
M(p^2) & = \displaystyle m_\mu \> + \> \left [M(p^2) \overline{ \Sigma}_d(p^2) \> + \>\overline{ \Sigma}_s(p^2)\right] \> -
\> \left[m_\mu \overline{ \Sigma}_d(\mu^2) \> + \> \overline{ \Sigma}_s(\mu^2)\right]\,, 
\label{eq:massfunc}   \\[3mm]
\overline{\Sigma}_s(p)&=  \frac{\alpha}{4\pi^3}\, \int_E d^4k\ \frac{1}{q^2}\,\frac{F_k}{\left[k^2+M_k^2\right]}\,\Bigg[ 
 \frac{\xi}{q^2}\,\frac{1}{F_p}\, \left[k \cdot q\, M_p-p \cdot q M_k \right]
\nonumber\\
&+  G_q \Bigg\{
                 \,\,\,\,\,\,\,\,\,\, \frac{1}{2}\, \left[ \frac{1}{F_k}+\frac{1}{F_p} \right] M_k \left[3  \right]   + \frac{1}{2(k^2-p^2)}\,\left[ \frac{1}{F_k} -\frac{1}{ F_p} \right] M_k \left[\frac{-4 \Delta^2}{q^2} \right] + \frac{1}{(k^2-p^2)}\,
\left[ \frac{M_k}{F_k} -\frac{M_p}{F_p} \right]\,\left[\frac{2\Delta^2}{q^2}\right]
\Bigg\}\,  \nonumber\\
& +  
 G_q \Bigg\{\,\,\,\,\,\,\,
                                          \tau_2^E(p^2,k^2,q^2) \left[-2 \Delta^2 \right] M_k  + \tau_3^E(p^2,k^2,q^2) \left[-3 q^2 \right] M_k 
   %
     + \tau_6^E(p^2,k^2,q^2) \left[-3 (p^2-k^2) \right]M_k 
     \Bigg\}\,\,
\quad \Bigg] \,.
\label{eqn:scalar_self_energy} 
\end{align}
\end{widetext}
We have represented arguments by subscripts for brevity $F_k = F(k^2)$.

\subsection{\label{subsec:PhotonDSE}Photon Wave-function Renormalization}
The renormalized photon DSE from Eq.~(\ref{eq:mainsdph}) is
\begin{align}
    \Delta^{-1}_{\mu\nu}(q) = Z_3(\mu)\,\left(\Delta^0_{\mu\nu}\right)^{-1}(q) + \,Z_1(\mu)\,{\overline{\Pi}}_{\mu\nu}(q)\,,
\label{eq:mainsdph2}
\end{align}
where ${\overline{\Pi}}_{\mu\nu}$ is the photon vacuum polarization or self-energy obtained by evaluating the
photon DSE diagram using the Feynman rules.

If we contract the photon self-energy with $q^\mu$~:
\begin{align}
q^\mu {\overline{\Pi}}_{\mu\nu}(q)&= i e^2 N_F \> {\rm Tr} \int_M \widetilde{dk}\> \gamma_\nu \>S(k) \>(q \cdot \Gamma(p,k)) \>  S(p)\,,
\end{align}
and use the WGTI 
\begin{align}
q^\mu {\overline{\Pi}}_{\mu\nu}(q)&=i e^2 N_F \> {\rm Tr} \int_M\widetilde{dk}\> \gamma_\nu \>\left(S(k) -  S(p)\right)\,,
\label{eq:qpimu}
\end{align}
where $\widetilde{dk} = d^4k/(2\pi)^4$ and $p=k-q$.

Using an analogous procedure to the fermion propagator, we can form the
appropriate subtractions of the renormalized photon DSEs, Eq.~(\ref{eq:mainsdph2}) to eliminate the divergent renormalization
constants $Z_1$ and $Z_3$ by recalling that $G(\mu^2;\mu^2) \>=\> 1$ yields~:
\begin{widetext}
\begin{align}
G^{-1}(\mu^2;q^2)\>& = \displaystyle \>1 \> + \> \left[ G^{-1}(\mu^2;q^2)\bar \Sigma_d(\mu^2) \> + \> \bar \Pi(q^2)\right] \> -
\>\left[ \bar \Sigma_d(\mu^2)\> + \> \bar \Pi(\mu^2) \right] \;\;\;,  
\label{eq:subphoton}  \\[3mm]
   \overline{\Pi}(q^2)
    &= \frac{\alpha \, N_F}{3\pi^3} \int_E \text{d}^4 k \frac{1}{q^2} \,
        \frac{F_p}{(p^2 + M_p^2)} \, \frac{F_k}{(k^2 + M_k^2)} \, \Bigg\{ \frac{1}{2} \left(\frac{1}{F_k} + \frac{1}{F_p}\right) \,
        \left[ 2 k \cdot p - \frac{8}{q^2}\left(\Delta^2+q^2 k \cdot p\right) \right] \nonumber\\[3mm]
&+ \frac{1}{2} \frac{\left(1/F_k - 1/F_p\right)}{(k^2-p^2)} \,
        \bigg[ \left(-\left(k^2+p^2\right) + 2 M_k M_p \right)
                \left\{\frac{8}{q^2} \left(k\cdot q\right)^2 - 3 k \cdot q - 2 k^2 \right\} -3 \left(k^2-p^2\right) \left(M_k M_p - k^2\right) \bigg] \nonumber \\[3mm]
&+\frac{ \left(M_k/F_k - M_p/F_p\right)} {k^2-p^2} \,
        \bigg[ -\left(M_k + M_p \right)
                \left\{\frac{8}{q^2} \left(k\cdot q\right)^2 - 3 k \cdot q - 2 k^2 \right\}  +\,3 \left(k^2-p^2\right) M_k \bigg] \,,            \nonumber  \\[5mm]
& + \tau_2^E(p^2,k^2,q^2) \,
    \left[ \left(k^2+p^2\right) \left\{-\Delta^2\right\} + M_k M_p \left\{2\Delta^2\right\} \right]+ \tau_3^E(p^2,k^2,q^2) \,
    \left[ 3 q^2 k \cdot p + 2\Delta^2 + M_k M_p \left\{3q^2\right\} \right] \nonumber \\
&+ \tau_6^E(p^2,k^2,q^2) \,
    \left[ 3k \cdot p \left(p^2 - k^2\right) + M_k M_p \left\{ 3\left(p^2-k^2\right)\right\} \right] + \tau_8^E(p^2,k^2,q^2) \,
    \left[ -2\Delta^2 \right] \quad  \Bigg\}\,\,.  
	\label{eqn:photon_self_energy}
\end{align}
\end{widetext}

\bibliography{literature}
\end{document}